%% file: main.tex
\documentclass{article}

\usepackage{arxiv}
\usepackage{lineno}
\usepackage[english]{babel}
\usepackage{color}
\usepackage{graphicx}
\usepackage{tabularx}
\usepackage{enumitem}
\usepackage[caption=false]{subfig} 
\usepackage{amsmath,amssymb,bm}
\usepackage[version=3]{mhchem}
\usepackage{verbatim}
\usepackage{multirow}
\usepackage{dcolumn}
\usepackage{mathtools}
\usepackage{float}
\usepackage{nicefrac}
\usepackage{siunitx}

\usepackage[utf8]{inputenc} % allow utf-8 input
\usepackage[T1]{fontenc}    % use 8-bit T1 fonts
\usepackage{hyperref}       % hyperlinks
\usepackage{url}            % simple URL typesetting
\usepackage{booktabs}       % professional-quality tables
\usepackage{amsfonts}       % blackboard math symbols
\usepackage{nicefrac}       % compact symbols for 1/2, etc.
\usepackage{microtype}      % microtypography
\usepackage{lipsum}		
\usepackage{graphicx}
\usepackage[numbers,sort&compress]{natbib}
\usepackage{doi}
\usepackage{setspace}
\usepackage{subfiles}

\renewcommand{\thefigure}{\arabic{figure}}
\renewcommand{\thetable}{\Roman{table}}

\makeatletter

\newcommand{\makesititle}{%
  \vbox{%
    \hsize\textwidth
    \linewidth\hsize
    \vskip 0.1in
    \@toptitlebar
    \centering
    {\LARGE\sc \@title\par}
    \@bottomtitlebar
    \textsc{\undertitle}\\
    \vskip 0.1in
    \def\And{%
      \end{tabular}\hfil\linebreak[0]\hfil%
      \begin{tabular}[t]{c}\bf\rule{\z@}{24\p@}\ignorespaces%
    }
    \def\AND{%
      \end{tabular}\hfil\linebreak[4]\hfil%
      \begin{tabular}[t]{c}\bf\rule{\z@}{24\p@}\ignorespaces%
    }
    \begin{tabular}[t]{c}\bf\rule{\z@}{24\p@}\@author\end{tabular}%
    \vskip 0.4in \@minus 0.1in
    \center{\@date}
    \vskip 0.2in
  }
}
\makeatother

\definecolor{orange}{rgb}{0.93, 0.53, 0.18}
\definecolor{bostonuniversityred}{rgb}{0.8, 0.0, 0.0}
\definecolor{darkspringgreen}{rgb}{0.09, 0.45, 0.27}

\title{%
Force-Free Molecular Dynamics Through Autoregressive Equivariant Networks 
}

\author{ \href{https://orcid.org/0000-0003-2951-6740}{\includegraphics[scale=0.06]{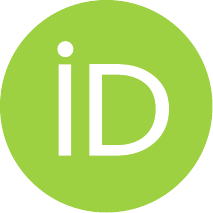}\hspace{1mm}Fabian L.~Thiemann}\thanks{To whom the correspondance should be addressed.}
        \\
	IBM Research Europe\\
	Daresbury, WA4 4AD, United Kingdom\\
	\texttt{fabian.thiemann@ibm.com} \\
	\And
	\href{https://orcid.org/0000-0002-9793-3450}{\includegraphics[scale=0.06]{orcid.pdf}\hspace{1mm}Thiago~Reschützegger} \\
	IBM Research\\
	Rio de Janeiro, 20031-170, RJ, Brazil\\
    \And \href{https://orcid.org/0009-0006-4159-568X}{\includegraphics[scale=0.06]{orcid.pdf}\hspace{1mm}Massimiliano~Esposito} \\
    	IBM Research Europe\\
	Daresbury, WA4 4AD, United Kingdom\\
    \And	\href{https://orcid.org/0000-0001-8847-7433}{\includegraphics[scale=0.06]{orcid.pdf}\hspace{1mm}Tseden~Taddese} \\
    The Hartree Centre \\
    Science and Technology Facilities Council\\
    Daresbury, WA4 4AD, United Kingdom\\
    \And \href{https://orcid.org/0000-0002-5384-3349}{\includegraphics[scale=0.06]{orcid.pdf}\hspace{1mm}Juan D.~Olarte-Plata} \\
    The Hartree Centre \\
    Science and Technology Facilities Council\\
    Daresbury, WA4 4AD, United Kingdom\\    
       \And \href{https://orcid.org/0000-0002-5350-8225}{\includegraphics[scale=0.06]
       {orcid.pdf}\hspace{1mm}Fausto~Martelli} \\
    	IBM Research Europe\\
	Daresbury, WA4 4AD, United Kingdom\\
}
\renewcommand{\undertitle}{}
\renewcommand{\headeright}{}

\hypersetup{
pdftitle={trajcast},
pdfsubject={-}
pdfauthor={-},
pdfkeywords={First keyword, Second keyword, More},
}

\begin{document}

\maketitle

\begin{abstract}
Molecular dynamics (MD) simulations play a crucial role in scientific research.
Yet their computational cost often limits the timescales and system sizes that can be explored. 
Most data-driven efforts have been focused on reducing the computational cost of accurate interatomic forces required for solving the equations of motion.
Despite their success, however, these machine learning interatomic potentials (MLIPs) are still bound to small time-steps.
In this work, we introduce \textit{TrajCast}, a transferable and data-efficient framework based on autoregressive equivariant message passing networks that directly updates atomic positions and velocities lifting the constraints imposed by traditional numerical integration.
We benchmark our framework across various systems, including a small molecule, crystalline material, and bulk liquid, demonstrating excellent agreement with reference MD simulations for structural, dynamical, and energetic properties.
Depending on the system, \textit{TrajCast} allows for forecast intervals up to $30\times$ larger than traditional MD time-steps, generating over 15 ns of trajectory data per day for a solid with more than 4,000 atoms.
By enabling efficient large-scale simulations over extended timescales, \textit{TrajCast} can accelerate materials discovery and explore physical phenomena beyond the reach of traditional simulations and experiments.
An open-source implementation of \textit{TrajCast} is accessible under \url{https://github.com/IBM/trajcast}.
\end{abstract}

% keywords can be removed

\section*{Introduction}

Molecular Dynamics (MD) simulations are an essential tool in modern science, providing mechanistic insight into complex phenomena, such as protein folding~\cite{levitt_computer_1975,lindorff-larsen_how_2011}, the effects of pressure on materials~\cite{dong_stable_2017,cavazzoni_superionic_1999,sun_phase_2015,rescigno_observation_2025}, the existence of liquid-liquid transitions in supercooled liquids~\cite{sastry_liquidliquid_2003,poole_phase_1992,palmer_metastable_2014}, and amorphization~\cite{rosu-finsen_medium-density_2023,cassone_electrofreezing_2024,fan_microscopic_2024}, to name but a few, that are difficult or impossible to study through experiments alone.
By numerically integrating the equations of motion for all particles in the molecule or material of interest, these simulations generate a trajectory, $\mathcal T \coloneq \left\{ \mathbf{x}(t)\right\}_{t=t_0}^{t=t_\mathrm{max}}$, where each state $\textbf{x}(t)$ is defined by the positions, $\mathbf{r}(t)\in \mathbb{R}^{3N}$, and velocities, $\mathbf{v}(t)\in \mathbb{R}^{3N}$, of the $N$ atoms at given time $t$.
However, obtaining trajectories at first principles accuracy is computationally intensive, typically limiting them to systems of a few hundred atoms and timescales up to one nanosecond.
These computational costs arise from (1) accurately approximating the Schrödinger equation to obtain reliable forces, and (2) the necessity for millions of integration steps due to the typically small time-step of 0.5 – 1.0 femtoseconds (fs).
Accordingly, two strategies have emerged to speed up trajectory generation, focusing on either improving force prediction efficiency or evolving the system with larger time-steps.\\

Machine learning interatomic potentials (MLIPs) fall into the first group and can accelerate the calculation of \textit{ab initio}-quality forces by orders of magnitude.
Trained on small systems where electronic structure calculations are feasible, these models operate locally, allowing for an accurate representation of the potential energy surface in broader, chemically similar systems without requiring direct reference data.
This has not only enabled access to much larger time and length scales, as demonstrated by the remarkable nanosecond-long simulation of the HIV virus with 44 million atoms~\cite{musaelian_scaling_2023}, but has also catalyzed exciting discoveries in simulations, including a new superionic phase of water under confinement~\cite{kapil_first-principles_2022} and structural and electronic phase transitions in silicon~\cite{deringer_origins_2021}.
The pursuit for faster and more accurate architectures has evolved from kernel-based methods to equivariant graph neural networks (GNNs)~\cite{behler_generalized_2007, bartok_gaussian_2010, schutt_schnet_2017, unke_physnet_2019,
drautz_atomic_2019, unke_spookynet_2021, batzner_e3-equivariant_2022, batatia_design_2025, batatia_mace_2022-1,musaelian_learning_2023, chmiela_accurate_2023,xie_ultra-fast_2023}, maturing into its own field~\cite{deringer_machine_2019, unke_machine_2021, behler_four_2021,thiemann_introduction_2025}.
This rapid progression has culminated recently in the introduction of several foundation models~\cite{chen_graph_2019,chen_universal_2022,merchant_scaling_2023,yang_mattersim_2024,batatia_foundation_2024} trained on large parts of the periodic table, which can be fine-tuned to the system of interest.\\

Multiple-time-step (MTS) integrators~\cite{tuckerman_molecular_1990,tuckerman_reversible_1992}, conversely, follow the second strategy aiming to loosen the constraint of small integration steps.
Specifically, they assign shorter time-steps to rapidly changing interactions, like atomic bonds, and longer time-steps to slower intermolecular interactions. 
In practice, MTS integrators are primarily used with classical potentials, which provide inherent scale separation through their analytical yet less accurate and transferable form.
A recent study~\cite{fu_learning_2023}, however, successfully coupled MTS with MLIPs by co-training a smaller model for short-time interactions, evaluated at each timestep during MD, alongside a larger, more expressive model which is called less frequently.
While this led to a $3\times$ speed-up, MTS integrators are fundamentally limited by the need for small time-steps in parts of the system, and further validation of scale-separated MLIPs for more complex systems is needed.\\

Generative machine learning approaches go beyond the idea of MTS, seeking to efficiently generate representative configurations or even complete trajectories without computing or integrating interatomic forces.
Boltzmann generators~\cite{noe_boltzmann_2019} directly approximate the Boltzmann distribution, producing states consistent with the equilibrium distribution at temperature $T$. However, these samples lack temporal order, obstructing insights into the system's temporal evolution and dynamical properties.
Recurrent neural networks have been shown to accurately interpolate MD trajectories of small molecules~\cite{winkler_high-fidelity_2022} and reliably predict the dynamics of 16 Lennard-Jones particles with time-steps over $1000 \times$ larger than in traditional MD simulations~\cite{kadupitiya_solving_2022}. These methods, however, operate with a fixed particle number, requiring retraining for unseen or larger systems, which restricts their applicability to realistic large-scale problems where accurate \textit{ab initio} reference data is often hard or impossible to obtain.\\ 

The flow-based \textit{Timewarp}~\cite{klein_timewarp_2023-1} model, conversely, generates MD trajectories via autoregressive roll-out with time-steps up to the order of $10^5$~fs and has been shown to generalize to unseen peptides.
Notwithstanding the speed-up of \textit{Timewarp} and its recent generative competitors, \textit{ITO}\cite{schreiner_implicit_2023-1} and \textit{MDGen}\cite{jing_generative_2024}, these approaches propagate atomic positions only, leaving the system's state incomplete without momenta.
Beyond impeding the computation of transport properties, this lack of full phase-space coverage introduces ambiguity;  the same atomic configuration can correspond to different velocity distributions and, thus, trajectories at different temperatures. This limits the ability of these methods to interpolate and extrapolate across thermodynamic variables.
Other deep learning architectures, such as the equivariant graph neural operator~\cite{xu_equivariant_2024} or the work presented by Fu \textit{et al.}~\cite{fu_simulate_2023}, incorporate assumptions about chemical bonds or restrict their focus to heavy atom motion. 
While these inductive biases may improve performance on specific tasks, they hinder generalizability to reactive systems where bond breaking and formation occurs.
Beyond architectural details, many of these methods require hundreds of nanoseconds of trajectory data to predict small molecule behavior accurately, making them less practical for scaling to complex systems.
Furthermore, nearly all generative models have been evaluated exclusively on small molecules and peptides, leaving their applicability to materials, condensed matter phases, and the accurate reproduction of equilibrium structural and dynamical properties largely unexplored.\\

In this work, we introduce \textit{TrajCast}, a  framework that employs equivariant message passing neural networks (MPNNs) to autoregressively predict the next state of a molecular system.
Unlike most MLIPs that derive forces from predicted energies, \textit{TrajCast} directly outputs the next state (positions and velocities) over a period $\Delta t$ which is at least one magnitude larger than the typical timestep in equivalent MD simulations.
By preserving the sequential nature of the trajectory, \textit{TrajCast} recovers the system's dynamics and enables the computation of time-dependent properties.
Generating full-state trajectories during roll-out ensures a complete system representation, enabling interpolation and extrapolation across different temperature regimes.
\textit{TrajCast} is free of assumptions about chemical bonds, making it capable, in principle, of modeling reactive systems.
Relying on local environments, \textit{TrajCast} is highly scalable and transferable, enabling models trained on small systems to be applied to larger or chemically similar ones.
Moreover, \textit{TrajCast} is very data-efficient, requiring significantly less than 1 ns of trajectory training data to achieve accurate predictions.
To our knowledge, \textit{TrajCast} is the only generative framework that integrates all these capabilities. 
We demonstrate its accuracy across a range of systems, from small molecules to crystalline solids and bulk liquids, highlighting its ability to consistently
reproduce structural, energetic, and dynamic properties, and showcasing its robustness and versatility across different domains.
Dependent on the system,
\textit{TrajCast} uses time intervals between $10\times$ and $30\times$ larger than traditional MD time-steps.
By avoiding gradient computations, this enables trajectory generation of more than 15 ns and 1 ns per day for crystalline quartz and liquid water, comprising 4,300 and 5,100 atoms, respectively.

% Enhanced sampling --> check timewarp for short sentence

% MD Gen -> conditional on first frame and tokenisation required for running an dnot clear for materials, no velocity

% Timewarp is transferable (train on amino acids and predict on others) -> not being tested for materials, non-physical velocities, temperature in boltzmann distribution though
% Coarse grained: coarse grained, prior knowledge of bonds, scalable?

% Boltzmann generatores sampling boltzmann distribution but not looking at the dynamics

% NVIDIA: Interesting but tests were performed without tracking hydrogens and not predicting velocities, only one step.

\section*{Results}

\subsection*{\textit{TrajCast} Framework}

\begin{figure*}[t]
    \centering
    \includegraphics[width=1\linewidth]{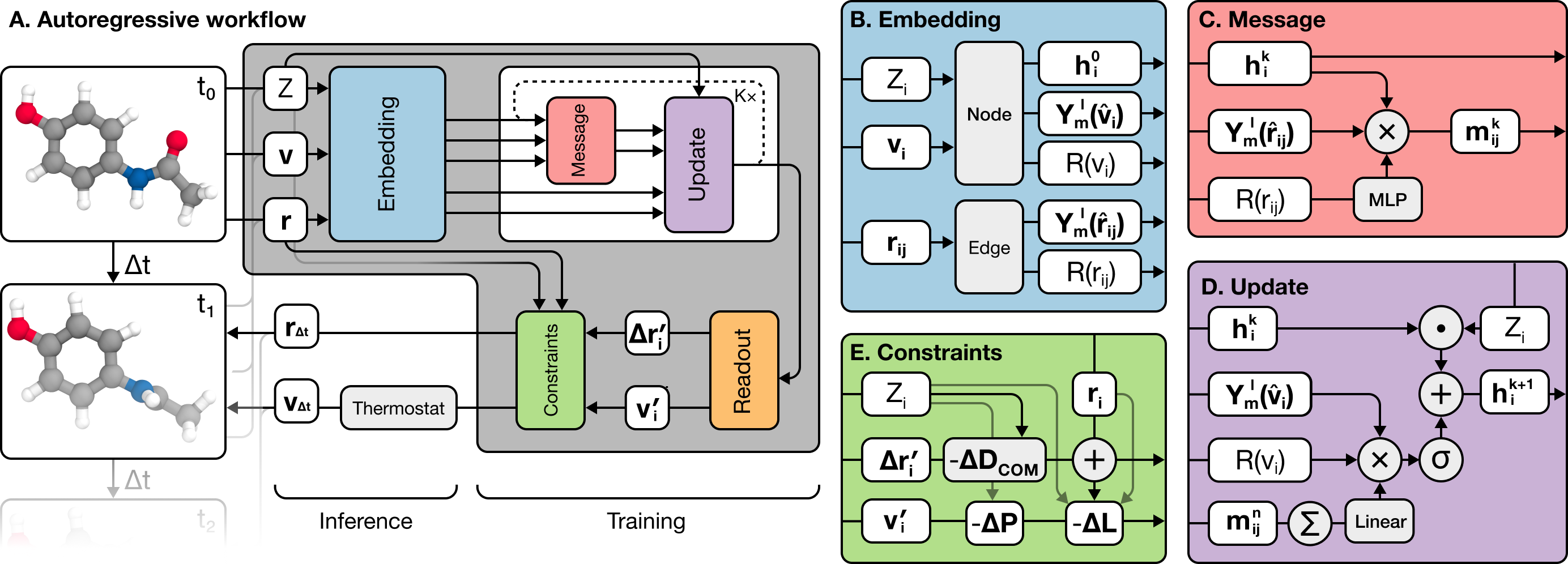}
    \caption{\textbf{Overview of the TrajCast architecture.}
    \textbf{(A)} Autoregressive workflow: An atomistic system at time $t_0$ is passed through an equivariant MPNN (grey box) to predict the new positions and velocities at time $t_1$. 
    Atomic attributes (positions, velocities, chemical elements) are encoded into initial features, which are refined over $T$ message passing blocks.
    Estimates of the displacement and velocity vectors are generated based on the final features.
    These are then refined to ensure momentum conservation.
    The trajectory is built by rolling out predictions, where outputs from one step serve as inputs for the next.
    A thermostat ensures sampling from the canonical (NVT) ensemble at constant temperature $T$, with states following the Boltzmann distribution.
    \textbf{(B)} The embedding block encodes node and edge attributes 
    %to express vectorial properties in a radial and spherical harmonics extension serving as the convolutional filters and p
    and generates the initial features.
    \textbf{(C)} Messages are constructed by convolving latent features with filters derived from a learnable radial basis and the spherical harmonics expansion of edge vectors.
    \textbf{(D)} In the update block, messages from neighbors are pooled and combined via a tensor product with velocity vectors in a learnable radial and spherical harmonic basis. The result is passed through a non-linearity and added to the previous layer's features, weighted by the node's chemical element.
    %
    % \textbf{(E)} The readout function estimates velocity and displacement vectors from the final node features ($l=1, p=-1$).
    %
    \textbf{(E)} Conservation of total linear and angular momentum is enforced by adjusting the displacements and velocities.
    }
    \label{fig:framework}
\end{figure*}

We start by outlining the \textit{TrajCast} architecture to predict temporal evolution of an atomistic system, as depicted in Fig.~\ref{fig:framework}.
\textit{TrajCast} produces trajectories autoregressively, where the state $\textbf{x}(t + \Delta t)$ is generated from $\textbf{x}(t)$ and used to predict the next state at the following time interval $\Delta t$ (Fig.~\ref{fig:framework}A), thus preserving the Markovian nature of traditional MD.
To enable comparison with experiments at constant temperature, we aim to sample trajectories from the canonical ensemble at constant temperature $T$.
We achieve this using a strategy inspired by the CSVR thermostat~\cite{bussi_canonical_2007}, where each time step is followed by a rescaling of the atomic velocities to enforce canonical sampling.
Rather than relying on forces to propagate the system using numerical integration in the microcanonical (NVE) ensemble with a small timestep $\delta t$, however, an equivariant MPNN predicts the updated positions and velocities at a larger timestep $\Delta t \gg \delta t$, as shown in the grey box in Fig.~\ref{fig:framework}A.
Additional model details are provided in the methods section, and here we focus on the key elements and deviations from previous work.\\

Compared to MLIPs, our equivariant MPNN takes the atomic velocities $\mathbf{v}(t)$ in addition to the positions $\mathbf{r}(t)$ and atomic types $Z$ as input.
In an embedding layer (Fig.~\ref{fig:framework}B) both node and edge attributes are encoded to produce the initial node features $\textbf{h}_i^0$ and express the edge and velocity vectors in a radial and spherical harmonics basis. 
For the chemical element $Z_i$ and the magnitude of the edge vectors ${r}_{ij}$, we follow the same approach as the MLIPs NequiP~\cite{batzner_e3-equivariant_2022} and MACE~\cite{batatia_mace_2022-1}, using a one-hot encoding and learnable Bessel functions multiplied with a polynomial cutoff as introduced originally in DimeNet~\cite{gasteiger_directional_2020}, respectively.
Since the atomic velocity depends on the particle mass, we encode its magnitude $v_i$ using a Gaussian basis set, where each Gaussian is weighted according to the chemical element of node $i$.
Thus, the $\mu$th element of the encoding is given by:
\begin{equation}
    R_\mu({v_i}) = \sum_{m=1}^M {W}_{Z_i,m} \phi_m (v_i)
    \label{eq:vel_enc}
\end{equation}
where $\phi_m$ is the $m$th Gaussian and $W_{Z_i,m}$ is a learnable weight dependent on atomic species $Z_i$.\\

The latent features are updated through a series of convolutions in the message passing layer highlighted by the white box in Fig.~\ref{fig:framework}A.
The message from node $j$ to node $i$ in layer $k$, schematically shown in Fig.~\ref{fig:framework}E, is constructed similarly to that in NequIP~\cite{batzner_e3-equivariant_2022}, and given by
\begin{equation}
        \textbf{m}_{ij}^k = \mathbf{h}_j^k \otimes_{\mathrm{CG}}^{\mathrm{MLP}^k_e(R(r_{ij}))} \mathbf{Y}(\hat{\textbf{r}}_{ij}) \mbox{ . }
    \end{equation}
Here, $\otimes_\mathrm{CG}^{\mathrm{MLP}^k_e(R(r_{ij}))}$ represents the Clebsch-Gordan (CG) tensor product, which generates all equivariant combinations of irreducible representations (irreps) of node features and the spherical harmonic embedding of the edge vector projected on the unit sphere, $\hat{\textbf{r}}_{ij}$.
The individual terms are then weighted by a multilayer perceptron (MLP), which take the radial basis embeddings of the edge vector magnitudes $r_{ij}$ as input.
For details on the derivation, the reader is referred to textbooks on representation theory overview and references ~\cite{thomas_tensor_2018}, \cite{brandstetter_geometric_2022}, or \cite{batatia_design_2025} for applications to GNNs and atomistic modeling.\\

The constructed pairwise messages, $\textbf{m}_{ij}$, only contain information about the topology of the neighborhood $\mathcal{N}(i)$.
To incorporate velocities in the feature update (Fig.~\ref{fig:framework}D), we follow an approach introduced by Brandstetter \textit{et al.}~\cite{brandstetter_geometric_2022}.
After an average pooling of the pairwise messages $\textbf{m}_{ij}$, we compute a node-based CG tensor-product between the aggregated message, $\textbf{m}_i$, and the spherical harmonic embedding of the velocity vector projected on the unit sphere, $\hat{\textbf{v}}_{i}$, resulting in
\begin{equation}
    \mathbf{\tilde{m}}^k_i = \mathbf{m}_i^k \otimes_{\mathrm{CG}}^{\mathrm{MLP}^k_v(R(v_{i}, Z_i))} \mathbf{Y} (\hat{\textbf{v}}_{i}) \mbox{ . }
\end{equation}
Similar to the first tensor product, the respective paths are weighted based on the species dependent embedding of the velocity magnitudes, $R(v_i, Z_i)$, passed through a MLP.
We constrain the dimensionality of $\mathbf{\tilde{m}}_i^k$ by applying a linear self-interaction after the average pooling step.
The new node features $\textbf{h}_i^{k+1}$ are then computed in a ResNet-like approach: 
The conditioned message $\mathbf{\tilde{m}}_i^k$ is passed through a non-linear gate~\cite{weiler_3d_2018} and added to the transformed node features from the previous layer, which are processed using an element-dependent linear self-interaction layer, as proposed in MACE~\cite{batatia_mace_2022-1, kovacs_evaluation_2023-1}.\\

After $K$ message passing layers, the vector components ($l=1, p=-1$) of the final latent features $\textbf{h}_i^{K}$ are passed through two linear layers to produce the initial estimate of the displacement vector $\mathbf{\Delta r}'_i$ and the new velocity $\textbf{v}'_i$.
Predicting relative displacements instead of absolute positions makes the model predictions independent of the dimensions of the simulation box.
Since both total linear and angular momentum are conserved in the NVE ensemble, the displacements and velocities are refined by removing any center of mass motion, $\mathbf{\Delta D}_{\mathrm{COM}}$, and excess momenta, $\mathbf{\Delta P}$ and $\mathbf{\Delta L}$, generated during the process as illustrated in Fig.~\ref{fig:framework}D.
The new positions are then calculated by adding the displacements to the current atomic positions, while the refined velocities are rescaled by a CSVR thermostat~\cite{bussi_canonical_2007} transitioning the system from the NVE ensemble to the NVT ensemble with positions $\textbf{r}_{\Delta_t}$ and velocities $\textbf{v}_{\Delta_t}$ that follow the Boltzmann distribution.
These outputs are used as inputs for the next step, gradually building the system's trajectory.
While \textit{TrajCast} can, in principle, operate without a thermostat to forecast a trajectory in the NVE ensemble, long autoregressive roll-outs tend to accumulate errors, leading to instability.
Beyond maintaining a constant temperature, in the course of this work, we found that the noise injected by the CSVR thermostat helps to stabilize the model's predictions, as previously observed in the context of turbulence modeling~\cite{stachenfeld_learned_2022}.\\

% no gradients
% locality -> scalability
% training:
% - NVE data -> run with feasible MD timestep and then displacements and velocities are computed over larger time interval $\Delta t$. These are the labels for the model. 
% - At the moment, the model infers its prediction horizon directly from the data. This requires the usage of a single timestep across the training set and obeys to train on a set of diverse prediction horizons.
% In principe we could run NVE simulaitons, however, autoregressive models are prone to error propagation which we seem to be mitigated by adding noise in terms of thermostat to the velocities whcih seem to stabilise the system
% exact loss function goes into training details in method

We train our equivariant MPNN on configurations generated during MD runs within the NVE ensemble using common time-steps of 0.5–1.0 fs. 
Relative displacements and new velocities are then computed over larger time intervals $\Delta t$.
During training, the model weights are adjusted to minimize the residual between the true and predicted displacement and velocity vectors.
Since the current version of \textit{TrajCast} infers the prediction horizon directly from the data, the training set needs to be precomputed at a fixed $\Delta t$.
Similar to traditional MD simulations, the choice of $\Delta t$ involves a trade-off:
Smaller values inherently result in higher accuracy, as the system becomes increasingly decorrelated over time, but are less efficient, requiring more forward passes to produce the desired trajectory.
By predicting atomic displacements and velocities based on the local environment encoded in latent features, our approach ensures transferability across different chemical spaces and scalability to much larger systems than those encountered during training. 
Unlike MLIP-driven MD, \textit{TrajCast} operates in force-free manner and does not rely on computing gradients with respect to the positions to propagate the system in time.
Our framework buildson the \textit{e3nn}~\cite{geiger_e3nn_2022} and \textit{cuEquivariance} Python libraries, which ensure the efficient equivariant transformation and propagation of geometric objects within deep learning frameworks.

\subsection*{Experiments}

We validate the accuracy of \textit{TrajCast} across various systems.
In addition to the small molecule paracetamol, we demonstrate its performance on crystalline $\alpha$-quartz and liquid water.
% %
For each system, we train several models across different prediction horizons $\Delta t$.
We report the mean absolute errors (MAE) for displacements and velocities for all MPNNs on a hold-out test set in section \ref{si:sec_maes} of the SI, showing that MAEs increase with larger $\Delta t$.
Besides deviating from the true NVE trajectory, large MAEs can cause instabilities, leading to unphysical behavior. 
Here, we showcase results for the largest $\Delta t$ that accurately reproduces the chosen properties and produces a stable trajectory.
Since our methodology samples from the canonical ensemble, direct comparison with MD is infeasible, as trajectories will inevitably depart due to the thermostat-induced randomness.
Instead, we assess whether ensemble averages of dynamic, structural, and energetic properties are accurately reproduced.
We demonstrate in section \ref{si:sec_nve} of the SI, however, that \textit{TrajCast} can reproduce a NVE trajectory accurately when run without a thermostat. %
Finally, we conclude our experiments, by comparing the data efficiency of \textit{TrajCast} across systems and prediction intervals $\Delta t$.

\begin{figure*}[t]
    \centering
    \includegraphics[width=0.9\linewidth]{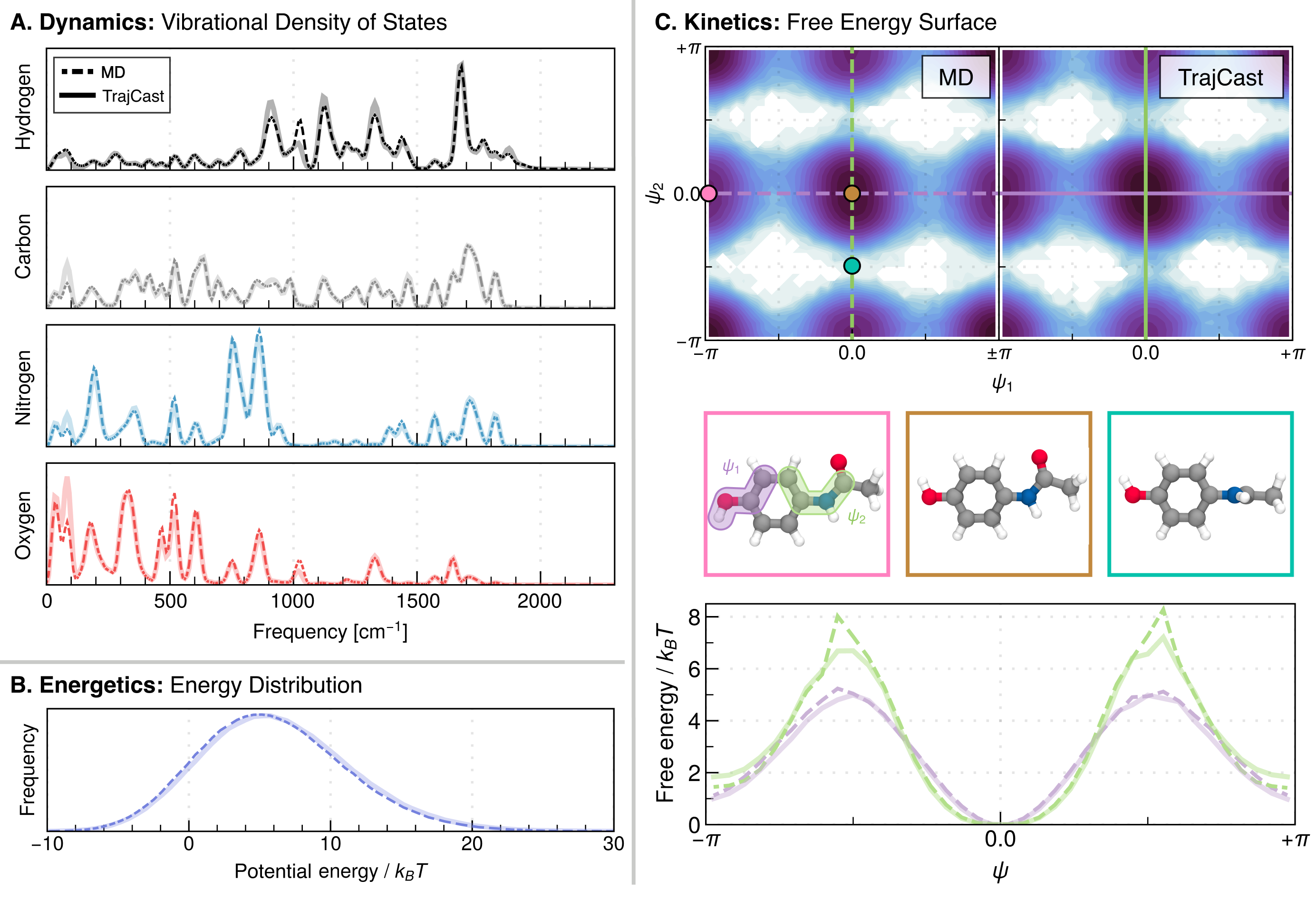}
    \caption{\textbf{Comparison of properties of paracetamol from MD and \textit{TrajCast} with a 7~fs prediction horizon.}
    (\textbf{A}) Dynamical properties are compared based on the element-specific vibrational density of states (VDOS).
    \textbf{(B)} Energetics are validated by comparing the distribution of the potential energy, computed in post-processing for the \textit{Trajcast}-generated trajectory by using the same force field as in the reference MD.
    (\textbf{C}) Free energy barriers drive the system’s kinetics and are computed from the free energy surface (FES), shown for both MD and \textit{TrajCast} at the top.
    The FES is generated by sampling two dihedral angles, $\psi_1$ and $\psi_2$, visualized by the inset within the bottom panel.
    The center panel visualizes different realizations of paracetamol for various dihedral pairs, matching the colors of corresponding points in the FES.
    Barriers are calculated along the cuts shown on the FES and colored accordingly in the bottom panel.
    }
    \label{fig:exps_paracetamol}
\end{figure*}

\subsubsection*{Small Molecules: Paracetamol}

We start by benchmarking the proposed methodology based on paracetamol, a small molecule included in the (rev)MD17 datasets~\cite{chmiela_machine_2017,christensen_role_2020} commonly used for validating MLIPs.
Paracetamol is chosen for its size and the presence of four distinct chemical elements (H, C, N, and O).
Rather than using the original datasets, we perform classical MD simulations and produce a set of 10,000 training configurations sampled from NVE trajectories equilibrated at $300$~K.
The prediction horizon is set to $\Delta t = 7~$fs, 14 times larger than the timestep used in the MD reference.
With the MPNN coupled to the CSVR thermostat, we generate a stable $7$~ns trajectory ($10^6$ steps) to compute ensemble averages and properties for comparison against MD NVT trajectories.
For a fair comparison, all properties from the MD reference were computed using the same frame frequency as the prediction horizon in \textit{TrajCast}.
Specifically, we validate our approach by assessing its accuracy in reproducing the molecule's dynamics, energetics, and kinetics, as shown in Fig.~\ref{fig:exps_paracetamol}.\\

For the dynamics, we analyse the species-resolved vibrational density of states (VDOS), which provides a comprehensive description of intramolecular and, if applicable, intermolecular motion.
Panel A of Fig.~\ref{fig:exps_paracetamol} displays the VDOS for all chemical species, computed from both MD and TrajCast trajectories.
With the chosen prediction horizon, the highest resolvable frequency is $\approx 2400~\mathrm{cm}^{-1}$, which prevents detecting intramolecular vibrations of bonds like C–H and O–H, with characteristic frequencies around or above $3000~\mathrm{cm}^{-1}$.
Within the resolvable range, our model shows excellent agreement with the MD reference, achieving overlap scores of $\approx 0.95$, following the method in reference~\cite{schran_machine_2021}.
%
% While MD using larger time-steps typically cause shifts in the VDOS peaks due to integration errors and aliasing, this artifact is absent in \textit{TrajCast}, where peaks align with the reference generated with a 14 times smaller time-step. This indicates that \textit{TrajCast} preserves the time-step of the training data without introducing prediction interval-related artifacts.
While larger time-steps in MD cause shifts in the VDOS peaks due to integration errors and aliasing, this artifact is absent in \textit{TrajCast}, where peaks align with those from the small MD time-step.\\
% %
% This suggests \textit{TrajCast} preserves the training time-step without introducing prediction horizon-related artifacts.\\

Next, we report the potential energy distribution in Fig.~\ref{fig:exps_paracetamol}B, computed from all configurations in both \textit{TrajCast} and MD trajectories using the same force field.
We find that the distributions match very closely, indicating that \textit{TrajCast} indeed samples from the Boltzmann distribution and produces a trajectory within the NVT ensemble.
This also provides evidence that structural features, such as bonds, angles, and dihedrals, are well recovered.
Building on this, we explore how well \textit{TrajCast} resolves the system's kinetics based on free energy barriers in panel C of Fig.~\ref{fig:exps_paracetamol}.
Specifically, we compute the free energy surface (FES) as a function of two dihedral angles, $\psi_1$ and $\psi_2$, corresponding to the torsional rotation of the hydroxyl and amine group, respectively, highlighted in the left box in center panel of Fig.~\ref{fig:exps_paracetamol}C.
The FES from both MD and TrajCast agree very well qualitatively, as shown in the top panel of Fig.~\ref{fig:exps_paracetamol}C. 
In both cases, paracetamol preserves planarity and adopts either a \textit{cis} or \textit{trans} configuration, with the hydroxyl and amine groups pointing in the same or opposite directions, respectively.
These configurations can be realized by different sets of dihedral angles related by symmetry: \textit{cis} configurations correspond to $|\psi_1| \leq \pi/2$ and $|\psi_2| \leq \pi/2$, or $|\psi_1| \geq \pi/2$ and $|\psi_2| \geq \pi/2$; \textit{trans} configurations occur when $|\psi_1| \geq \pi/2$ and $|\psi_2| \leq \pi/2$, or $|\psi_1| \leq \pi/2$ and $|\psi_2| \geq \pi/2$.
Snapshots of paracetamol corresponding to selected dihedral pairs in the FES are shown in the central panel of Fig.~\ref{fig:exps_paracetamol}C.
For a more quantitative comparison, we compute the barriers along cuts at $\psi_1 = 0$ and $\psi_2 = 0$, shown in the bottom panel of Fig.~\ref{fig:exps_paracetamol}C. 
\textit{TrajCast} closely matches the shape and barrier heights predicted by MD, with nearly identical free energy profiles for hydroxyl dihedral rotation ($\psi_2 = 0$) and deviations within $\approx  k_B T$ for varying $\psi_2$, peaking at the local maxima ($\psi_2 \to \pm \pi/2$).
Overal, the agreement in kinetics, energetics, structural characteristics, and dynamics provides strong evidence that TrajCast can reliably predict accurate trajectories of isolated molecules.

\begin{figure*}[t]
    \centering
    \includegraphics[width=0.6\linewidth]{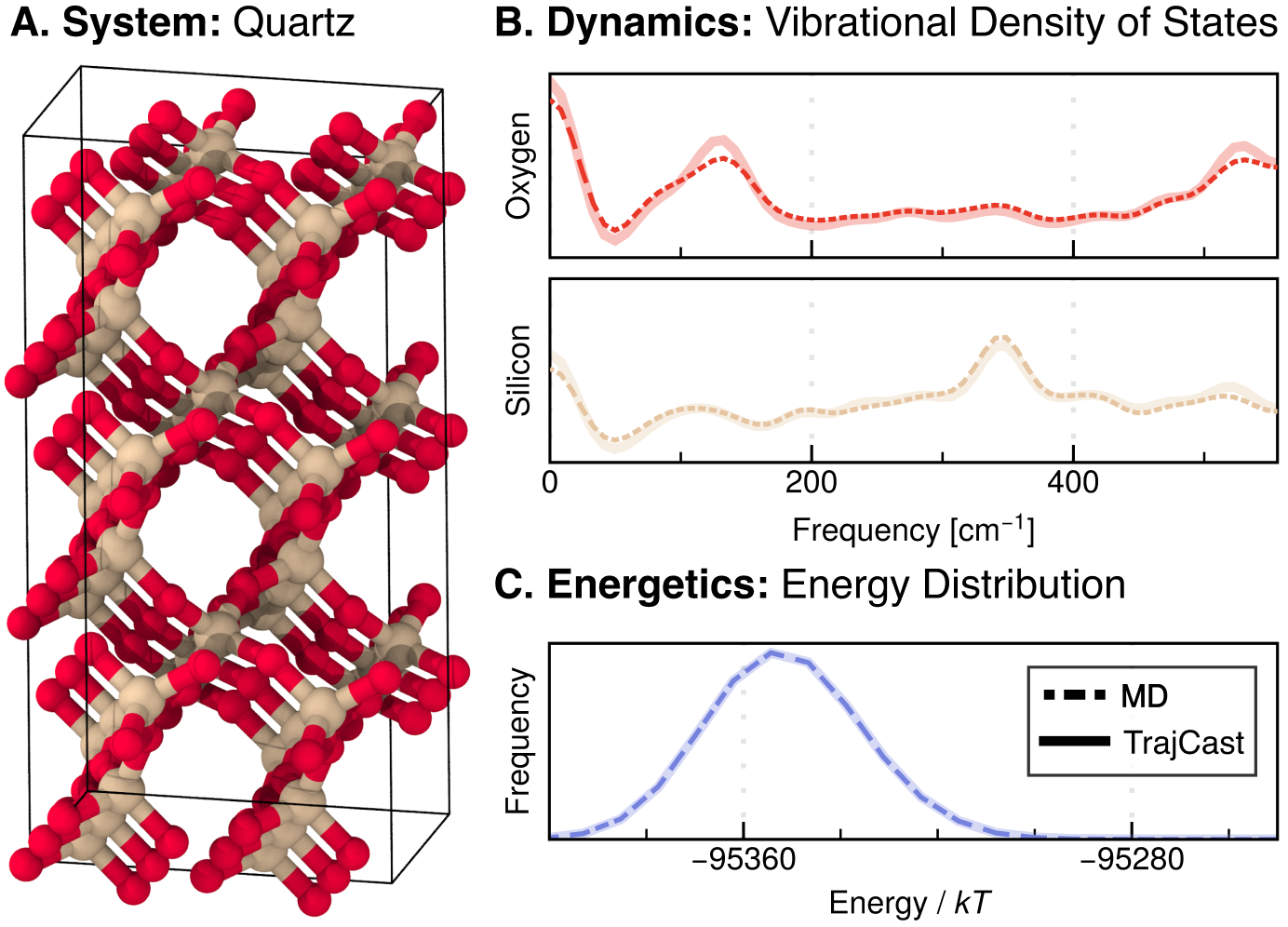}
    \caption{
    \textbf{Evaluation of \textit{TrajCast} on crystalline quartz with a 30~fs prediction horizon.}
    (\textbf{A}) Illustration of an atomic configuration of quartz used to train and validate \textit{TrajCast}.    
    (\textbf{B}) Dynamical properties are compared based on the element-specific vibrational density of states (VDOS).
    % %
    \textbf{(C)} Energetics are validated by comparing the distribution of the potential energy, computed in post-processing for the \textit{Trajcast}-generated trajectory by using the same force field as in the reference MD.
    }
    \label{fig:exps_quartz}
\end{figure*}

\subsubsection*{Crystalline Solid: Quartz}

We now move to benchmarking our framework for condensed matter phases, beginning with crystalline $\alpha$-quartz illustrated in  Fig.~\ref{fig:exps_quartz}A.
To this end, we train our MPNN on 5,000 samples from MD simulations of systems comprising 162 atoms at 300~K. 
In contrast to atoms of gas-phase molecules, particles forming a periodic lattice exhibit a more constrained and correlated motion, which allows us to set the prediction horizon to $\Delta t = 30$~fs, a 30-fold increase over the MD timestep, while still generating stable and accurate trajectories.
Similar to the previous section, we compare \textit{TrajCast} and MD based on the VDOS and energy distribution as shown in Fig.~\ref{fig:exps_quartz}B and ~\ref{fig:exps_quartz}C, respectively.
These properties are calculated from trajectories of 15~ns length corresponding to 500,000 iterative \textit{TrajCast} predictions.
We find that \textit{TrajCast} agrees very well with the MD reference for both quantities achieving VDOS overlap scores~\cite{schran_machine_2021} above 0.95 for both atom types within the resolvable frequency range and almost identical energy distribution.
The peak at zero frequency is an artifact caused by the small system size and low time-sampling resolution. 
We observe that it disappears in both MD and \textit{TrajCast} when the trajectory is sampled at intervals of 10~fs or lower.
The accurate reproduction of dynamics, energetics, and structural characteristics at large time intervals suggests \textit{TrajCast} as a useful tool for studying equilibrium properties of solids.
In section \ref{si:sec_efficiency} of the SI, we demonstrate that our approach scales efficiently with system size, enabling the generation of nearly 20~ns per day for a $3\times3\times3$ quartz supercell comprising about 4,300 atoms.
As a comparison, MD simulations using the widely used MLIP MACE architecture typically produce around 1~ns of trajectory per day for 1,000 atoms~\cite{batatia_foundation_2024}.
Beyond pristine lattices, this could make \textit{TrajCast} well-suited for exploring defective materials, where studying phenomena at low defect concentrations requires particularly large system sizes.

\begin{figure*}[t]
    \centering
    \includegraphics[width=0.9\linewidth]{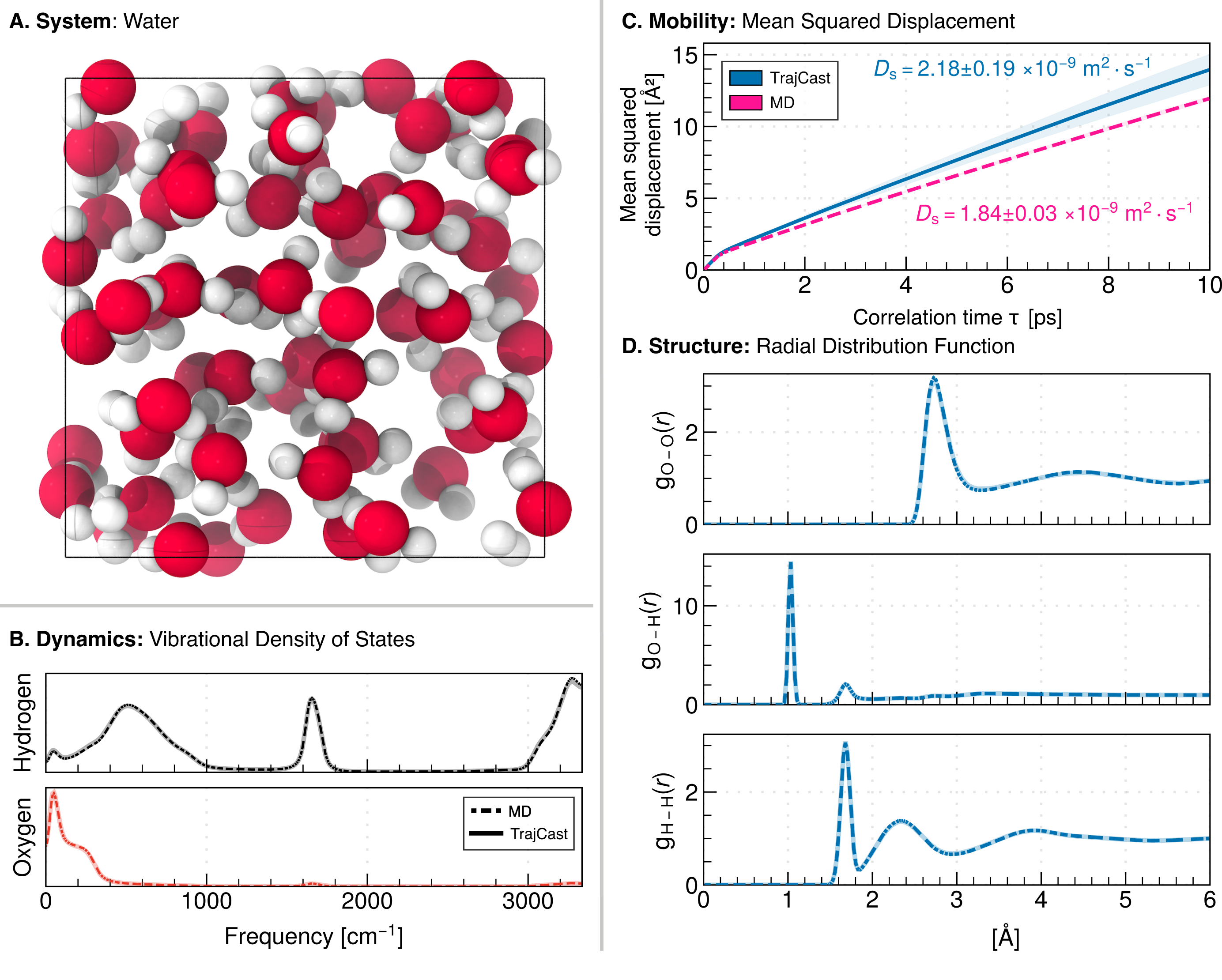}
    \caption{
    \textbf{Benchmarking \textit{TrajCast} on liquid water with a 5~fs prediction horizon.}
    (\textbf{A}) Snapshot of an instantaneous configuration of bulk water.     
    (\textbf{B}) Comparison of the element-specific vibrational density of states (VDOS) between \textit{TrajCast} and MD.
    % %
    \textbf{(C)} Mean squared displacement (MSD) of oxygen atoms over a correlation time $\tau$ of up to 10~ps, complementing the dynamic evaluation.
    (\textbf{D}) Validation of the molecular liquid structure based on the pairwise combinations of chemical elements in the radial distribution function (RDF).
    The shaded areas represent the threefold standard deviation, computed over 4 blocks.
    The diffusion coefficients are computed based on a fit to the slope in the range from 1 to 10~ps following the procedure outlined in~\cite{marsalek_quantum_2017}.
    }
    \label{fig:exps_water}
\end{figure*}

\subsubsection*{Liquid: Bulk Water}

As a final test case, we focus on liquid bulk water, which requires an accurate description of both intramolecular and intermolecular interactions.
Specifically, we use a system of 64 water molecules at ambient pressure and 300~K, a common setup for ab initio studies of bulk water~\cite{marsalek_quantum_2017} illustrated in Fig.~\ref{fig:exps_water}A, with a training budget of 5,000 samples.
We report results for a prediction horizon of $\Delta t = 5$~fs, as this provides the best trade-off between stability and accuracy.
While ten times larger than the typical MD timestep, it remains smaller than the prediction horizons explored for $\alpha$-quartz, reflecting the more complex motion of a molecular liquid compared to a crystalline solid.
To evaluate the accuracy of our model, we generate a 2.5~ns trajectory, corresponding to 500,000 roll-out steps, and compare it with a MD trajectory of the same length.
In addition to the VDOS, we additionally validate \textit{TrajCast} based on the mean squared displacement (MSD) and the radial distribution function (RDF).
The former provides insight into the molecular mobility, whereas the RDF enables comparison of the local structure across all pairwise combinations of chemical elements.\\

Starting with the VDOS in Fig.~\ref{fig:exps_water}B, TrajCast nearly perfectly reproduces the spectrum for both oxygen and hydrogen throughout the entire frequency range, achieving overlap scores~\cite{schran_machine_2021} of $\approx 0.99$.
This strongly suggests that our framework accurately captures molecular motion across different modes, from high-frequency intramolecular bending and stretching to low-frequency intermolecular translation and libration.
To complement the validation of the dynamics from a position-based perspective, we next examine the mobility of water molecules based on the MSD for a correlation time $\tau$ of up to $10$~ps visualized in Fig.~\ref{fig:exps_water}C.
While \textit{TrajCast} achieves excellent agreement with its MD reference in the ballistic regime, $\tau \lessapprox 0.5$~ps, MSDs diverge slightly at longer timescales which is likely due to error accumulation during the roll-out process.
To quantify these deviations, we compute the diffusion coefficient from the slope of the MSDs in the diffusive regime $\tau\geq 1.0$~ps obtaining $2.18\pm0.19$ and $1.84\pm0.03$ $\times 10^{-9}$ m$^2$ s$^{-1}$, for \textit{TrajCast} and MD respectively.
Although \textit{TrajCast} predicts slightly more mobile water, the diffusion coefficients are in close agreement, indicating that our framework can reliably and consistently predict the dynamics of molecular liquids.
Beyond the dynamics, we analyze the liquid's structure by comparing the RDF, finding excellent agreement across all three pairwise combinations of chemical elements shown in Fig.~\ref{fig:exps_water}D. 
The only slight deviation occurs in the oxygen-oxygen RDF, where \textit{TrajCast} produces a slightly understructured liquid, consistent with the marginally higher diffusion coefficient.\\

While these benchmarks suggest that our methodology can accurately capture the dynamical and structural characteristics of bulk liquids at room temperature, we believe \textit{TrajCast} could be valuable for studying systems dominated by slow dynamics, where long but challenging simulations are required~\cite{angell_formation_1995,debenedetti_supercooled_2001,scalliet_thirty_2022}.
For deeply undercooled liquids, for instance, these conditions may enable larger prediction horizons, though accurately accounting for cooperative effects and the emergence of spatial and temporal heterogeneities remains essential~\cite{stokely_effect_2010}.
Even with $\Delta t = 5.0$~fs, however, \textit{TrajCast} generates over 1 ns of simulation for a $3\times3\times3$ supercell with more than 5,000 atoms per day as highlighted in section \ref{si:sec_efficiency} of the SI.

\begin{figure*}[t]
    \centering
    \includegraphics[width=1\linewidth]{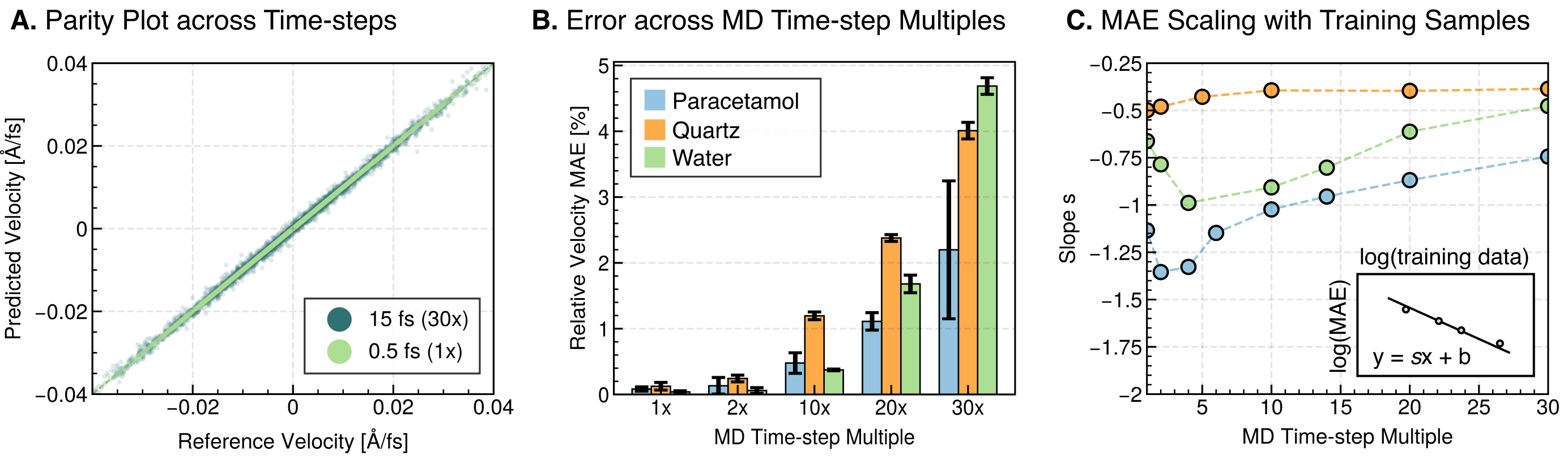}
    \caption{
    \textbf{Impact of prediction horizon and training set size on performance of \textit{TrajCast}.}
    % %
    (\textbf{A}) Parity plot comparing predicted and reference velocities for \textit{TrajCast} models trained for water with prediction horizons of 0.5 and 15~fs, respectively.
    (\textbf{B}) Dependence of the relative velocity MAE on the MD time-step multiple.
    Error bars correspond to the threefold standard deviations computed across three models varying in the initial weights.
    (\textbf{C}) Exponents from power law fits to learning curves, quantifying the relationship between model accuracy and training set size across different chemical systems and MD time-step multiples.
    The inset shows that these exponents correspond to the slope $s$ of a linear fit
    on a log-log plot of MAE versus training set size.
    The larger the absolute value of $s$ the larger the improvement on the generalization error of the model when additional data is used.
    The dashed lines serve as guide to the eye.
    % % %
    % \textbf{(C)} Mean squared displacement (MSD) of oxygen atoms over a correlation time $\tau$ of up to 10~ps, complementing the dynamic evaluation.
    % %
    % (\textbf{D}) Validation of the molecular liquid structure based on the pairwise combinations of chemical elements in the radial distribution function (RDF).
    % %
    % The shaded areas correspond to the statistical error computed from averaging over 4 blocks. 
    % %
    % The diffusion coefficients are computed based on a fit to the slope in the range from 1 to 10~ps following the procedure outlined in \cite{marsalek_quantum_2017}.
    }
    \label{fig:scalability}
\end{figure*}

\subsubsection*{Scalability and Data Efficiency}
So far, we have presented results for a single prediction horizon and a fixed number of training samples, both of which are expected to significantly impact model performance. 
To explore this, we conduct additional experiments varying both factors for all three systems and summarize the results in Fig~\ref{fig:scalability}.
For each experiment, we train three identical models varying only the randomly assigned initial weights. 
To facilitate comparison across systems, we express the prediction horizon as a multiple of the original MD timestep (MTM)  and focus on MAEs computed over a test set rather than specific properties.
We specifically choose the velocity MAE, as velocities remain consistent across different prediction horizons within each system.\\

We first describe results from models trained on varying prediction horizons, using the same number of training samples as in previous experiments.
While scientists have developed an intuition for acceptable force errors in MLIPs over the past decade, this understanding must first be established for velocity errors in \textit{TrajCast}.
To this end, we present a parity plot for models trained on MTMs of 1 and 30 for water in Fig.~\ref{fig:scalability}A and achieving a velocity MAE of $\approx 3\times10^{-6}$~\AA$/\mathrm{fs}$ and $\approx 4\times 10^{-4}$~\AA$/\mathrm{fs}$, respectively.
These correspond to relative errors of $\approx 0.03\%$ and $\approx 4.64\%$ when expressed with respect to the mean absolute velocity.
Interestingly, the parity plot visually suggests two well-performing models despite their MAE differing by over two orders of magnitude. 
However, during roll-out, none of the three models with an MTM of 30 was able to produce a stable trajectory longer than a few ten-thousand steps.
While MLIPs can tolerate relative force errors of up to $30\%$~\cite{ying_advances_2025}, this indicates that \textit{TrajCast} requires significantly smaller relative errors to generate stable and physical trajectories.\\

To further explore the sensitivity with respect to the prediction horizon across the investigated systems, Fig.~\ref{fig:scalability}B shows the relative velocity MAE for \textit{TrajCast} for MTMs of 1, 2, 10, 20, and 30.
In sections \ref{si:sec_maes} and \ref{si:sec_learning} of the SI, we report the MAE for additional MTMs as well as the scaling of the displacement MAE, which follows a similar trend but with consistently smaller relative errors by about 40\%.
For all systems, we find that the error increases with MTM with water exhibiting exponential scaling and the steepest increase.
While the molecular motion is highly correlated at small time intervals, resulting in small errors, the system quickly decorrelates at larger MTMs due to the stochastic motion of the fluid.
It also becomes clear that the maximum acceptable model error depends on the specific system.
For instance, our model for crystalline quartz at a MTM of 30 produces stable trajectories despite a relative velocity MAE of approximately $3.96\%$, whereas a \textit{TrajCast} model for water at a MTM of 20 and a relative velocity MAE of $1.62\%$ may produce instabilities.\\

Arguably, a simple solution to stabilize large MTM models and speed up trajectory generation is to train on larger datasets.
To assess the impact of training set size on model performance across chemical systems and MTMs, we train multiple TrajCast models with identical architecture on $10\%$ and $50\%$ of the dataset size used in the experiments above.
In section \ref{si:sec_learning} of the SI, we show the MAE for both displacements and velocities as a function of dataset size across all systems and MTMs.
These learning curves typically follow a power law~\cite{hestness_deep_2017}, where the exponent $s$ determines how quickly model performance improves with more data.
%
% In a log-log plot, this exponent corresponds to the slope of a straight line as illustrated in schematic inset in Fig.~\ref{fig:scalability}C.
%
In Fig.~\ref{fig:scalability}C, we show $s$ as a function of MTM for all systems, with a schematic inset illustrating that $s$ corresponds to the slope of a straight line in a log-log plot.
Interestingly, the dependence of $s$ on MTM varies significantly across chemical space.
While the curve for quartz increases monotonically from $\approx-  0.5$ to $\approx -0.4$, paracetamol and bulk water exhibit a non-monotonic relationship, reaching a minimum at $\mathrm{MTM}\approx 4$ before increasing to smaller absolute values.
Naturally, learning the system's evolution over larger time intervals is more difficult, requiring more data to capture the variety of potential new states and reducing the absolute value of $s$.\\

Across the range of MTMs investigated, paracetamol consistently shows the largest slope, followed by water and quartz, reflecting the varying constraints in each system:
The atomic motion of quartz is coupled to its rigid lattice, while water molecules can move more freely but remain strongly correlated with neighboring molecules. In contrast, gas-phase paracetamol is solely governed by intramolecular interactions, requiring fewer examples to resolve, even at long time intervals.
These observations align with learning curve experiments with MLIPs, where $s$ is roughly $3\times$ larger for MACE on an organic molecule~\cite{batatia_mace_2022-1} compared to NequIP for water~\cite{batzner_e3-equivariant_2022}.
Notably, their largest slopes are up to 4 times smaller than those obtained for our models on similar systems, suggesting that incorporating velocities extracts more information from each configuration, improving data efficiency.
This enables training accurate \textit{TrajCast} models on dataset sizes similar to MLIPs, using just a few hundred picoseconds of trajectory data instead of the typical hundreds of nanoseconds required by other methods~\cite{kadupitiya_solving_2022,fu_simulate_2023, klein_timewarp_2023-1, jing_generative_2024}
This is crucial as generating large datasets at \textit{ab initio} quality becomes infeasible due to the high computational cost, especially when scaling to multiple systems or building foundation models.

% MACE \cite{batatia_mace_2022} and NequIP \cite{batzner_e3-equivariant_2022}
% compare the values to MLIP 
% say why this is veyr important as data generation is expensive and we do not want to train on as many trajectories as XXX

% To this end, we generate learning curves for TrajCast across all systems and various MTMs, where each learning curve corresponds to a log-log plot of predictive error as a function of dataset size, as shown in the SI and schematically illustrated in the inset of Fig.~5C.

\section*{Conclusion}

We have introduced \textit{Trajcast}, a new methodology based on autoregressive equivariant MPNNs to accelerate the generation of accurate MD trajectories.
Instead of predicting forces for numerical integration, our framework directly updates velocities and positions, bypassing the need for small integration steps.
We have benchmarked our methodology on a variety of chemical systems ranging from a small molecule to condensed matter phases such as crystalline material and bulk liquid.
 In all cases, \textit{TrajCast} reliably reproduces structural, dynamical, and energetic properties, while enabling $10 \times$ to $30 \times$ larger time steps than traditional atomistic MD simulations.
 Importantly, \textit{TrajCast} is transferable and scales well to larger systems, allowing to generate over 1 ns of trajectory data per day for 1,728 water molecules and more than 15 ns for a quartz supercell with more than 4,300 atoms.
In contrast, the current MACE MLIP architecture, enables MD simulations of about 1~ns for 1,000 atoms per day~\cite{batatia_foundation_2024}.
 Notably, our models are trained using only a few hundred ps of trajectory data, a quantity feasible to generate with AIMD.
 The current version of \textit{TrajCast} operates with fixed forecast horizon, however, future work will explore multi-time-step training to which could improve generalizibility and accuracy over longer timescales, as demonstrated by a recent autoregressive GNN for weather forecasting~\cite{lam_learning_2023}. 
 The current force-free approach of \textit{TrajCast} prevents the computation of pressure and other force-related properties, restricting forecasting to the microcanonical (NVE) and canonical (NVT) ensembles. 
 Extending the methodology to the isobaric-isothermal (NpT) ensemble, however, will be an interesting avenue for future work.
Additionally, although we benchmarked our methodology on various systems beyond isolated molecules, it would be interesting to explore how it performs when trained on multiple regions of chemical space simultaneously—an important step toward developing a foundation model for molecular dynamics.
While \textit{TrajCast} has the potential to accelerate atomistic simulations and material discovery in general, we believe a particularly promising application lies exploring slow physical phenomena at low temperatures.
In these systems, slow relaxation times not only require long simulation runs—as seen in a recent study~\cite{sciortino_constraints_2025} on supercooled water that took several years to reach the microsecond scale—but also impose higher correlations, potentially enabling to leverage larger forecasting horizons. 
Thus, \textit{TrajCast} could greatly accelerate this process, making large-scale simulations significantly more efficient.
By addressing the long-standing challenge of accessing long simulation times, our work could help bridge currently disjoint length scales—a key goal in many scientific disciplines.

\section*{Methods}

\subsection*{Equivariant Message Passing Neural Networks for Atomistic Systems}
MPNNs, a subclass of GNNs, learn mappings from molecular graphs to target properties such as the potential energy in MLIPs.
In these graphs, nodes represent atoms, connected by edges usually determined through a distance criterion.
Each node $i$ has attributes like its atomic type $Z_i$ and position~$\textbf{r}_i$ in three-dimensional (3D) Euclidean space.
Depending on the task, additional attributes may include the velocity~$\textbf{v}_i$ or the force~$\textbf{f}_i$.
Edge attributes typically consist of the Euclidean vectors $\textbf{r}_{ij}$ connecting two nodes $i$ and $j$.\\

The local state of each node is encoded in latent features $\textbf{h}_i^t$ which are iteratively updated over $T$ message passing layers.
In each layer $t$, nodes exchange information via messages $\textbf{m}^t_{ij}$, aggregrated from all neighbors and used to update the node's latent state.
Formally, this is expressed as~\cite{gilmer_neural_2017}:
\begin{align}
    \textbf{m}^t_i &= \sum_{j \in \mathcal{N}(i)} \textbf{m}_{ij}^t =\sum_{j \in \mathcal{N}(i)} M^t(\textbf{h}_i^t, \textbf{h}_j^t, \textbf{e}_{ij})\\
    \textbf{h}^{t+1}_i &= U^t(\textbf{h}_i^t, \textbf{m}^t_i) \label{eq:mpnn_update}
\end{align}
where $\mathcal{N}(i)$ represents the set of nodes connected to node $i$ and $M^t$ and $U^t$ are arbitrary message and update functions for layer $t$.
While the original formulation omits node attributes in the update, advanced MLIPs like BOTNet~\cite{batatia_design_2025} or MACE~\cite{batatia_mace_2022-1, kovacs_evaluation_2023-1} and other MPNN architectures like SEGNNs~\cite{brandstetter_geometric_2022}, incorporate them to prevent information loss across layers.
After $T$ message passing steps, the final features may be pooled for global property prediction or passed through a readout function directly for node properties.\\

The latent features $\textbf{h}_i^t$ can consist of various types of geometric objects, such as scalars, vectors or higher order tensors, each transforming uniquely the Euclidean group $E(3)$ operations of translation, rotation, and inversion.
In practice, however, it suffices to focus on the orthogonal group $O(3)$ (rotations and reflections), as translational invariance is ensured by using interatomic vectors rather than absolute positions.
Transformations under $O(3)$ are described by irreducible representation (irreps) which are labelled by their rotation order $l= 0, 1, 2,\ldots$ and parity $p \in (1,- 1)$.
Scalars ($l=0, p=1$) remain invariant under rotation, while vectors ($l=1, p=-1$) transform consistently (equivariantly) with the geometry.
If the features only depend on the atomic positions, formally each element of the feature vector with rotation order $L$, $\textbf{h}_{i, L}^t$, is equivariant under any rotation $g = \textbf{R} \in O(3)$ if 
\begin{equation}
    \textbf{h}^t_{i,L}(\textbf{R} \cdot (\textbf{r}_1, \ldots, \textbf{r}_N)) = \textbf{D}_L(g) \textbf{h}^t_{i,L}(\textbf{r}_1, \ldots, \textbf{r}_N)  \quad  \forall g \in O(3) 
\end{equation}
where $\textbf{R}$ is the 3D rotation matrix parametrised by $g$ and $\textbf{D}_L(g)$ is the Wigner D-matrix representing the equivalent transformation in feature space. 
These symmetries can, in principle, be learned from augmented datasets comprising rotated configurations.
Alternatively, they can be explicitly incorporated into MPNNs by using equivariant message and update functions, ensuring the network respects rotational symmetry by construction.
The latter has become common practice in state-of-the-art MLIPs~\cite{batatia_design_2025, batatia_mace_2022-1, batzner_e3-equivariant_2022, musaelian_learning_2023}, where leveraging higher-order features beyond scalars has been shown to significantly enhance accuracy.

\subsection*{Ensembles and Thermostats}

By numerically integrating the equations of motion, MD simulations produce a trajectory that follows the distribution of the microcanonical (NVE) ensemble, with a constant number of particles $N$, volume $V$, and total energy $E$.
In this ensemble, both the total linear momentum $\textbf{P}$ and angular momentum $\textbf{L}$ are conserved.
However, since experiments are usually conducted at constant temperature rather than energy, MD simulations are coupled to a heat bath via a thermostat to sample from the canonical (NVT) ensemble, where the temperature $T$ fluctuates around the desired value.
Microscopically, the temperature $T$ of a $N$ particle system is defined by
\begin{equation}
    T = 2\frac{K}{N_f k_B} \quad \text{with} \quad K = \frac{1}{2} \sum_{i=1}^N m_i |\textbf{v}_i|^2 
    % \frac{1}{N_f k_B}\sum_{i=1}^N m_i |\textbf{v}_i|^2 
\end{equation}
where $K$ is the kinetic energy, $m_i$ is the mass of particle $i$, $k_B$ is the Boltzmann constant, and $N_f$ corresponds to the degrees of freedom, typically given as $3N-3$ for a 3D periodic system.\\

Velocity rescaling is the simplest method for adjusting a system to a target temperature, where atomic velocities are rescaled at a predefined frequency by a factor $\alpha = \sqrt{\hat{K}/K}$, with $\hat{K}$ representing the target kinetic energy.
However, this method has several drawbacks, including failure to sample from the canonical ensemble, generation of discontinuities, and violation of equipartition~\cite{braun_anomalous_2018}.
The canonical sampling through velocity rescaling (CSVR) thermostat~\cite{bussi_canonical_2007} overcomes these limitations by sampling $\hat{K}$ from a canonical equilibrium distribution and rescaling the velocities after each integration step.
The coupling strength, defined by the relaxation time $\tau$ relative to the integration timestep, must be chosen appropriately to maintain the target temperature without significantly altering the system's dynamics.

\subsection*{Reference Data Sets}

All data used for training and validation have been generated using classical MD simulations with the LAMMPS~\cite{plimpton_fast_1995,thompson_lammps_2022} software package.
Training trajectories were generated in the NVE ensemble, while data to assess the roll-out performance of the model architecture stems from NVT runs.
To ensure that the NVE trajectories are at the desired temperature, we first equilibrate the system in the NVT ensemble.
During the equilibration period, the total energy is sampled, and configurations with a total energy close to the distribution's average are subsequently selected.
The NVE runs are then initiated from five configurations that are uniformly sampled from this subset, ensuring that the selected snapshots are at least 1~ps apart.
The specific simulation settings for each system are detailed below.

\subsubsection*{Paracetamol}

An isolated paracetamol molecule is simulated in vacuum using the OPLS all-atom force field~\cite{jorgensen_opls_1988,robertson_improved_2015} instantiated with Moltemplate~\cite{jewett_moltemplate_2021}.
We use a time-step of 0.5~fs due to the presence of hydrogen atoms. 
The system is first equilibrated in the NVT ensemble for 25 ps (50,000 steps) using a Nosé-Hoover chain thermostat~\cite{martyna_nosehoover_1992}, with 10 chains, a damping factor of 5 fs, and a target temperature of 300 K.
This equilibration protocol is applied uniformly before subsequent production runs in both the NVE and NVT ensembles.
The NVE production runs are initiated from configurations where the total energy is within a relative threshold of 0.001 from the mean of the distribution.
For the NVT production run, we utilise the CSVR thermostat~\cite{bussi_canonical_2007}, consistent with the setup in \textit{TrajCast}, with a relaxation time of 50 fs. 
The temperature is calculated based on $3N-6$ degrees of freedom.
Net angular momentum is removed at each time-step during NVT simulations, while in the NVE ensemble, this quantity is naturally conserved, maintaining an initial value of $\textbf{0}$.
Production simulations are conducted for 100 ps (200,000 steps) in the NVE ensemble to generate training data, and for 10 ns (20,000,000 steps) in the NVT ensemble for validation.

\subsubsection*{Quartz}

Bulk $\alpha$-quartz containing 162 atoms is simulated using the van Beest, Kramer, and Santen (BKS) potential~\cite{van_beest_force_1990} with truncated Coulomb interactions~\cite{carre_amorphous_2007} and a timestep of 1~fs. 
%
%Velocities were initially generated randomly with a Gaussian distribution.
%
The system is equilibrated to 300~K in two stages: first in the NpT ensemble at ambient pressure using the Berendsen barostat~\cite{berendsen_molecular_1984} for 10~ps (10,000 steps) with a relaxation time of 0.1~ps, then in the NVT ensemble for 50~ps (50,000 steps) with Nos\'e-Hoover chain thermostat~\cite{martyna_nosehoover_1992} with 3 chains and a relaxation time of 0.1~ps.
Production runs of 100~ps (100,000 steps) in the NVE ensemble are initiated from five configurations randomly sampled from the equilibrated NVT trajectory, selecting those within a $\pm$10~K range of the target temperature.
The minimum time separation between selected configurations is set to 0.1~ps.
NVT simulations for comparison with \textit{TrajCast} predictions are carried out using the CSVR thermostat~\cite{bussi_canonical_2007} with a relaxation time of 100~fs and are performed for 15~ns (15,000,000 steps).

\subsubsection*{Liquid Water}

Bulk water is modeled using a system of 64 water molecules in a cubic simulation box with a fixed length of 12.4172 \AA. 
Intramolecular and intermolecular interactions of water is described by the flexible version of the simple point charge (SPC) interaction potential~\cite{toukan_molecular-dynamics_1985, wu_flexible_2006, smirnov_molecular_2016}. 
All MD simulations are carried out with a time-step of 0.5~fs. 
The system is equilibrated for 10~ps (20,000 steps) in the NVT ensemble using a Nos\'e-Hoover thermostat~\cite{martyna_nosehoover_1992} with a relaxation time of 0.5~ps and a target temperature of $300$~K.
The temperature is calculated based on $3N-3$ degrees of freedom.
Production runs of 50~ps (100,000 steps) in the NVE ensemble are initiated from five configurations randomly sampled from the equilibrated trajectory, selecting those within a $\pm$20~K range of the target temperature.
The minimum time separation between selected configurations is set to 0.1~ps.
Simulations in the NVT ensemble are carried out using the CSVR thermostat~\cite{bussi_canonical_2007} with a relaxation time of 50~fs and are performed for 2.5~ns (5,000,000 steps).

\subsection*{Model Size and Hyperparameters}

All equivariant MPNNs in this work use both even ($p=1$) and odd $(p=-1)$ parity features up to an rotation order of $l_{\mathrm{max}} =2$ creating a feature vector scalars, vectors, and higher-order tensors.
We use 64 channels per object, resulting in a feature vector described as a direct sum of irreps and expressed in \textit{e3nn}~\cite{geiger_e3nn_2022} notation as \texttt{64 x 0o + 64 x 0e + 64 x 1o + 64 x 1e +64 x 2o + 64 x 2e}, where each term specifies the multiplicity (first number), rotation order $l$, and parity $p$ ($o$ for odd, $e$ for even). 
For all experiments, we employ 4 message passing layers and use SiLU and tanh as activation functions in the non-linear gates~\cite{weiler_3d_2018} for even and odd invariants, respectively.\\

The \textit{TrajCast} MPNN contains approximately 2.2M learnable weights, most of which reside in the MLPs that process the radial embeddings of edge and velocity magnitudes. 
Each MLP comprises three hidden layers with 64 neurons per layer and uses SiLU activation functions.
Identical to NequIP~\cite{batzner_e3-equivariant_2022}, for the radial embedding of the edge lengths we use 8 learnable Bessel functions and a polynomial cutoff with $p=6$~\cite{gasteiger_directional_2020}.
Similarly, 8 Gaussian basis functions are used to generate the encoding of atomic velocity magnitudes.
The element-specific velocity encoding from equation \ref{eq:vel_enc} is concatenated with the chemical species embeddings of each node to form the initial features passed to the first message passing layer.\\

The cutoff radius for defining edges between nodes varies by system and is set to $4.0$~\AA, $4.5$~\AA, and $6.0$~$\text{\AA}$ for paracetamol, quartz, and liquid water, respectively.
The upper bound of the radial embedding of the velocities is determined by the lightest element, $Z_\mathrm{min}$ (hydrogen for water and paracetamol, oxygen for quartz), which governs the maximum velocity. 
For this element, we calculate the theoretical expectation value of the RMS velocity, $v_{\mathrm{RMS},Z_{\mathrm{min}}}^{\mathrm{MB}}$, based on the Maxwell-Boltzmann distribution.
When training across multiple temperatures, we recommend computing the RMS velocity for the highest temperature.
We found that the upper limit of the encoding works then best when set to values between $3.5\times v_{\mathrm{RMS},Z_{\mathrm{min}}}^{\mathrm{MB}}$ and $5 \times v_{\mathrm{RMS},Z_{\mathrm{min}}}^{\mathrm{MB}}$.\\

In all models, we refine the velocities predicted by the readout to remove any excess total linear momentum.
While angular momentum can be easily removed for isolated molecules such as paracetamol, a proper treatment is needed for periodic systems like bulk quartz or liquid water, where particles crossing boundaries and interacting with periodic images can introduce discontinuities and torques.
A common approach is to treat the simulation box as an open system, as suggested by Kuzkin~\cite{kuzkin_angular_2015}.
However, for bulk systems investigated here, angular momentum conservation is typically not a significant concern compared to more complex systems such as interfaces or multiphase systems, due to their isotropic nature and symmetric forces. 
In fact, we find that all MPNNs perform well even without additional constraints, with no significant excess angular momentum generated during the roll-out.

\subsection*{Training}

All models are trained based on the loss function
\begin{equation}
    \mathcal{L} = \frac{1}{3\sum_{b=1}^{N_b}N_a}\sum_{b=1}^{N_b}\sum_{i=1}^{N_a} \lambda_d \left(\widehat{\Delta r}_{i,\alpha}- \Delta r_{i,\alpha}\right)^2 + \lambda_v \left(\hat{ v}_{i,\alpha}- v_{i,\alpha}\right)^2 
\end{equation}
where $N_b$ denotes the batch size, $N_a$ is the number of atoms in each configuration within the batch, $\widehat{\Delta r}_{i,\alpha}$ and $\widehat{\Delta v}_{i,\alpha}$ represent the predicted $\alpha$-component of the displacement and velocity of atom $i$, respectively, and  $\lambda_d$ and $\lambda_v$ control the relative weighting of the displacement and velocity mean squared errors (MSE).
For all experiments and independent of the chosen prediction horizon of the model, we assign equal weights $\lambda_d=\lambda_v= 0.5$.
The target vectors are normalized by their root mean square (RMS) computed over the entire training set, and the model’s predictions are rescaled to match the correct magnitude.\\

The weights are optimized with the Adam optimizer~\cite{kingma_adam_2017} using AMSGrad~\cite{reddi_convergence_2019} and default values for $\beta_1=0.9$ and $\beta_2=0.999$.
The  number of training samples varies across experiments, with models trained on 1,000, 5,000, and 10,000 configurations of paracetamol, and 500, 2,500, and 5,000 snapshots of quartz and liquid water.
The results reported in the experiments correspond to models trained on the maximum sample size, unless stated otherwise.
All training configurations are drawn randomly from three independent NVE trajectories of each system.
The validation set comprises a quarter of the training set size, and a fixed hold-out test set of 1,000 configurations is used to compute final accuracy metrics.
Both validation and test set are sampled from independent trajectories, ensuring no overlap with the training set.\\ 

Irrespective of the training set size, a batch size of 10 for molecules and 2 for the condensed matter systems is used. 
The initial learning rate of 0.01 is adjusted by an on-plateau scheduler that reduced the rate by a factor of 0.8 if the validation loss does not improve over 20 epochs.
Training continues for a maximum of 1,500 epochs, with early stopping if no validation loss reduction is observed over 100 epochs.
Hyperparameters for the optimizer, learning rate, and batch size are optimized for the best balance between training time and model performance on the largest training set and prediction horizons.
Every experiment is performed using double precision. 
Both the model and training routine are implemented using the PyTorch library~\cite{paszke_pytorch_2019}.
All models are trained on a single NVIDIA V100-SXM2-32GB GPU, with training durations of up to 5, 10, and 11 days for paracetamol, quartz, and water, respectively, many completing earlier due to early stopping.

\subsection*{Trajectory Forecasting}

All trajectories generated using \textit{TrajCast} are produced autoregressively by rolling out the predictions of the MPNN, with velocities being rescaled after each step using a CSVR thermostat~\cite{bussi_canonical_2007}. 
As in classical MD simulations, the relaxation time of the thermostat, $\tau$, plays a crucial role and must be carefully chosen.
In section \ref{si:sec_thermo} of the SI, we explore how \textit{TrajCast} behaves with different values of $\tau$ for all systems and varying $\Delta t$. 
We find that for larger prediction horizons, much lower relaxation times are required compared to typical MD values to maintain the target temperature and fluctuations.
This is likely due to \textit{TrajCast} implicitly performing multiple timesteps before rescaling, whereas in MD, velocities are rescaled after each step.
For the reported experiments, we set $\tau$ to $10\times\Delta t$ (70~fs) for paracetamol, $2.5\times \Delta t$ (75~fs) for $\alpha$-quartz, and $10\times \Delta t$ (50~fs) for liquid water.
The target temperature for all systems is $300$~K and is defined based on $3N-6$ and $3N-3$ degrees of freedom for paracetamol and the condensed matter systems, respectively.
Initial velocities are sampled from a Gaussian distribution at $300~K$ with zero net linear and angular momentum.
For paracetamol, the net angular momentum is zeroed every 100 forward passes to correct for the excess angular momentum induced by the CSVR thermostat.
For paracetamol we produce a trajectory of 7 ns (1,000,000 forward passes), while for water and quartz, the roll-out is performed over 500,000 iterations, corresponding to 2 ns and 15 ns, respectively. 
All systems are equilibrated for $10\%$ of the total number of forward passes used for production.
In addition to PyTorch~\cite{paszke_pytorch_2019}, \textit{e3nn}~\cite{geiger_e3nn_2022}, and \textit{cuEquivariance}, we also use the ASE~\cite{hjorth_larsen_atomic_2017} python package to efficiently generate and save the trajectories.

\section*{Data and Code Availability}

An open-source implementation of \textit{TrajCast} is available at \url{https://github.com/IBM/trajcast} for training models and performing trajectory forecasting.
The model weights used in the experiments on various chemical systems are available at \url{https://huggingface.co/ibm-research/trajcast.models-arxiv2025}.
The training, validation, and testing datasets for these experiments can be accessed at \url{https://huggingface.co/datasets/ibm-research/trajcast.datasets-arxiv2025}.

\section*{Acknowledgements}
The authors thank Benjamin I. Tan, Christoph Schran, Clyde Fare, Edward O. Pyzer-Knapp, Eldad Haber, Johannes Brandstetter, Lior Horesh, Ljiljana Stojanovic, and Will Trojak for fruitful discussions. This work was supported by the Hartree National Centre for Digital Innovation, a collaboration between the Science and Technology Facilities Council and IBM.

\section*{Author Contributions}

F.L.T. designed research; F.L.T., T.R. and M.E. performed experiments; F.L.T., T.R., and M.E. wrote the code; F.L.T., J.D.O.-P., T.T., and F.M. generated data;  all authors analyzed data; and all authors wrote the paper.

\section*{Competing Interests}
The authors declare no competing interest.
%----------------------------------------------------------------------
\bibliographystyle{unsrt}
\bibliography{bibliography} 
%----------------------------------------------------------------------

\appendix
\clearpage
\setcounter{page}{1}
\setcounter{equation}{0}
\setcounter{figure}{0}
\renewcommand{\thepage}{S\arabic{page}} 
\renewcommand{\thetable}{S\arabic{table}}  
\renewcommand{\thefigure}{S\arabic{figure}}
\renewcommand{\thesection}{\Alph{section}}
\renewcommand{\thesubsection}{\thesection.\arabic{subsection}}
% \makesititle

\include{si_end}

\end{document}

%% file: si_end.tex
\section*{Supplementary Information}

In this supplementary information, we provide further details on specific aspects of the study presented in the manuscript. This includes additional metrics to evaluate the proposed methodology, as well as various aspects related to inference and generating new trajectories.

\tableofcontents
\newpage

\section{Model Performance}
In this section, we provide additional details supporting the validation of our framework as presented in the manuscript.
Specifically, we include an overview of the mean absolute error (MAE) across all systems for message passing neural networks (MPNNs) trained on the largest pool of training configurations.
Furthermore, we present additional results of the scalability and learning curve experiments to examine the impact of the prediction horizon and data efficiency.
% \section{TrajCast architecture}

\subsection{Mean Absolute Errors for Displacements and Velocities}\label{si:sec_maes}

Here, we report the MAE for displacements and velocities predicted by the equivariant MPNNs for all systems investigated in the manuscript for a range of prediction horizons $\Delta t$.
Detailed results are provided in Table \ref{tab:mae_all}.
The MAEs are calculated using $\Delta t$-specific hold-out test sets, each comprising $1,000$ atomic snapshots sampled from NVE trajectories not used for training or validation.
In addition to the absolute errors, we also report them relative to the mean absolute values of the displacement and velocity vectors. 
This is important, as the average magnitude of the displacement vector is strongly dependent on the chosen prediction interval $\Delta t$.
While we performed experiments with varying training set sizes and three different seeds, here we report the best models obtained with the largest number of training training samples.

\begin{table}[h!]
    \caption{\textbf{Displacement and velocity MAE for the systems studied at varying prediction horizons $\Delta t$.}
    Models for paracetamol were trained on a dataset of $10,000$ atomic structures, whereas models for $\alpha$-quartz and liquid water had a training budget of $5,000$ samples each.
    The MAEs are computed for hold-out test sets with varying $\Delta t$ comprising $1,000$ configurations, respectively.
    Beyond absolute values in $[\text{\AA}]$ and $[\text{\AA}/\mathrm{fs}]$, we also report the errors relative to mean absolute displacement and velocity of the test set.
    }
    \label{tab:mae_all}
    \resizebox{\textwidth}{!}{ % make sure the table is centered
        \begin{tabular}{lccccc}
             \toprule
             \midrule
             % &&&\\
             \multirow{2}{*}{System} & \multirow{2}{*}{Prediction horizon $\Delta t$ [fs]}& \multicolumn{2}{c}{Displacement} &  \multicolumn{2}{c}{Velocity}\\
             & & Abs [$10^{-5} \times \text{\AA}$]  & Rel [$\%$] &  Abs [$10^{-5} \times \text{\AA}/\mathrm{fs}$]& Rel [$\%$]
             \\ 
             \midrule
             \midrule
         \multirow{8}{*}{Paracetamol}&0.5 & 0.15 & 0.04& 0.52 & 0.07 \\
         &1.0 & 0.41 & 0.06& 0.71 & 0.10\\
         &2.0 & 1.56 & 0.11 & 1.42 & 0.20\\
         & 3.0 & 3.59 & 0.17 & 2.12 & 0.29\\
         & 5.0 & 7.98 & 0.24 & 3.05 & 0.42\\
        & 7.0 & 15.89 & 0.38 & 4.92 & 0.68\\
        & 10.0 & 31.91 & 0.59 & 7.88 & 1.08\\
        & 15.0 & 76.07 & 1.09 & 14.07 & 1.93\\
          \midrule   
         \multirow{7}{*}{Quartz} &1.0 & 0.16 & 0.06 & 0.32 & 0.11 \\
         & 2.0 & 0.66 & 0.11 & 0.64 & 0.22 \\
         & 5.0 & 4.71 & 0.33 & 1.80 & 0.63 \\
         & 10.0 & 18.51 & 0.69 & 3.37& 1.18\\
         & 20.0 & 66.90 & 1.50 & 6.77 & 2.36 \\
         & 30.0 & 141.24 & 2.52 & 11.37 & 3.96\\
         & 50.0 & 380.39& 5.51 & 25.13 & 8.76\\
        \midrule
        \multirow{8}{*}{Water} & 0.5 & 0.08 & 0.02 & 0.31 & 0.03\\
        & 1.0 & 0.24 & 0.03 & 0.45 & 0.05\\
        & 2.0 & 0.91 & 0.05 & 0.88 & 0.09\\
        & 5.0 & 6.99 & 0.17 & 3.46 & 0.37 \\
        & 7.0 & 15.38 & 0.29 & 6.38 & 0.68 \\
        & 10.0 & 47.23 & 0.67 & 15.25 & 1.63 \\
        & 15.0 & 199.04 & 2.04 & 43.47 & 4.64\\
        & 20.0 & 507.98 & 4.30 & 84.37 & 9.11\\
         
         \midrule
            \bottomrule
        \end{tabular}
    }
\end{table}

\subsection{Learning Curves Experiments}\label{si:sec_learning}

\begin{figure*}[t!]
    \centering
    \includegraphics[width=1\linewidth]{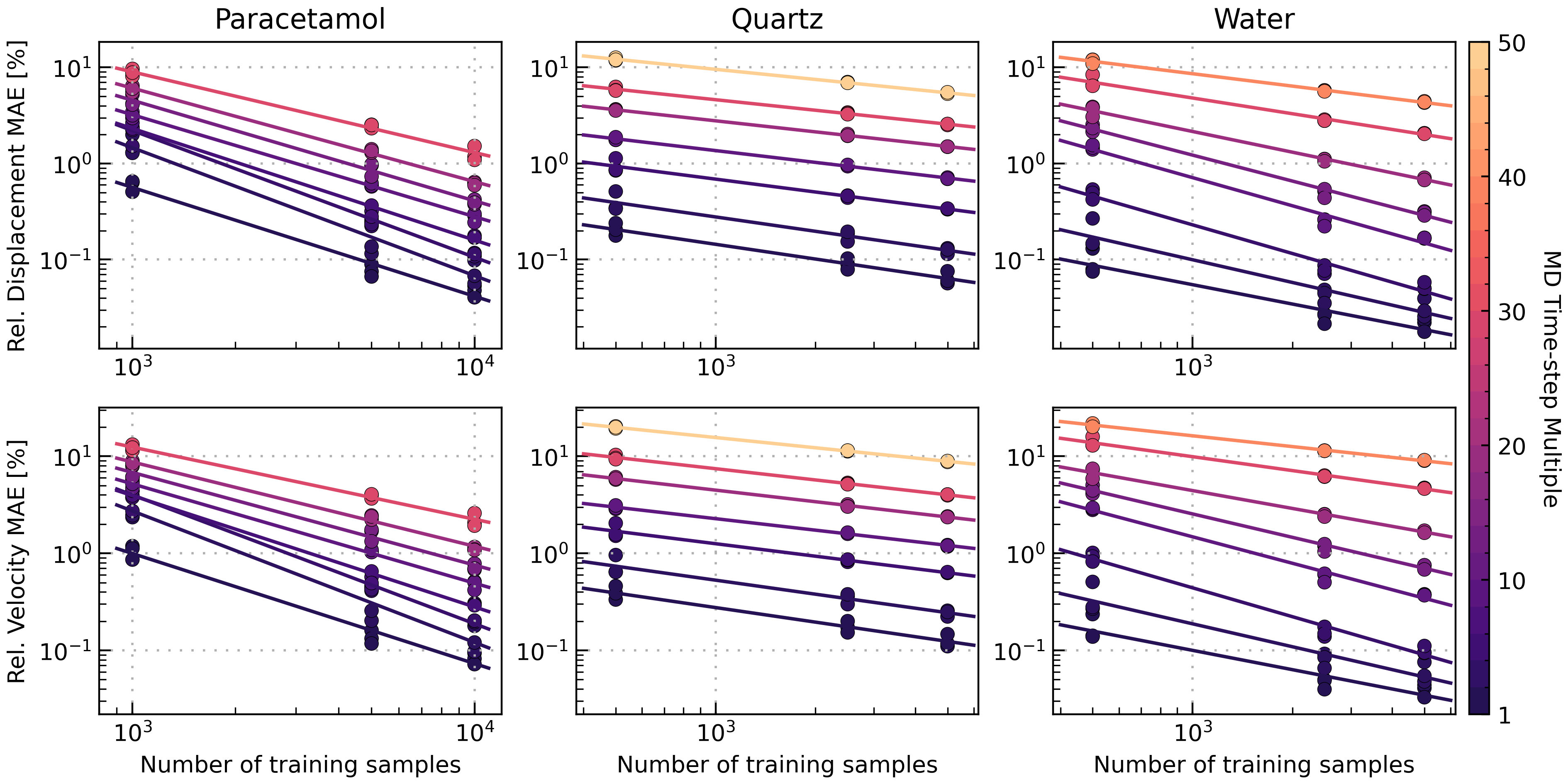}
    \caption{\textbf{Learning curves capturing the relationship between model accuracy and training set size across chemical systems and MD time-step multiples.}
    Top panels show the displacement MAE for paracetamol, quartz, and water, while the bottom panel displays the velocity MAE, both on a log-log scale.
    The MAEs were computed over a test set of 1,000 configurations and normalized by the mean of the displacements and velocities respectively.
    For each MTM and number of training samples we show the error for three models of identical architecture only differing in their initial weights and order they were fed with training samples.
    The continuous lines are fits to a power law to determine how fast the models learn when presented with additional data.
    }
    \label{si:fig_learning}
\end{figure*}

In this section we provide a full overview of the learning curves for both displacements and velocities.
The relative displacement and velocity MAEs for all systems and investigated MD time-step multiples (MTM) are shown as a function of number of training samples in Fig.~ \ref{si:fig_learning}.
As in the manuscript, the MAEs are normalized by the mean of the respective property sampled over the test set comprising 1,000 configurations.
Models were trained to 1,000, 5,000, and 10,000 or 500, 2,500, and 5,000 configurations for paracetamol and the condensed matter systems quartz and water, respectively.
For each combination of system, MTM, and number of training samples, we conducted three training runs to also study variation across different seeds.
We then fitted a power law \cite{hestness_deep_2017} $\mathrm{MAE} \propto N^s$, where $N$ is the number of training configurations and $s$ corresponds to the exponent quantifying how much the model improves with more data, to the generated data which appear as straight lines with slope $s$ in the log-log plot.
This allows us to compare data efficiency across systems and prediction horizons.
In the manuscript, we already discussed differences in the scaling behavior across systems and MTMs based on the velocity MAE.
The displacement MAE shows almost the identical relationship with respect to the number of training data suggesting a very similar exponent $s$.
We note, however, that the displacement MAE are consistently shifted to smaller MAEs across MTMs and number of training samples for all systems.
While we assigned equal weights to velocities and displacements in the loss function, this suggests opportunities for further optimization in future work.

\clearpage

\section{Trajectory Forecasting}

While the previous section focused on single-step prediction accuracy, here we provide additional details related to the autoregressive roll-outs and trajectory generation.
Specifically, we investigate the impact of thermostat strength over different prediction horizons, demonstrate the generation of stable trajectories in NVE for short periods, and benchmark \textit{TrajCast} inference timings across systems.

\subsection{Thermostat Relaxation Time} \label{si:sec_thermo}

The relaxation time $\tau$ of a thermostat is a crucial parameter in classical MD simulations.
If chosen too high, the thermostat cannot maintain the target temperature; if too low, it becomes overly strong, disrupting the system's dynamics and fluctuations.
Typically, $\tau$ is expressed relative to the integration time-step $\delta t$ and is chosen to be $\approx100-1000 \times \delta t$. 
However, it is unclear if the same rule of thumb applies to the methodology introduced in the manuscript.
While the CSVR thermostat \cite{bussi_canonical_2007} is designed to rescale velocities after each numerical integration step within the NVE ensemble, our trained MPNNs implicitly perform multiple NVE steps before rescaling the velocities.
Here, we perform a comprehensive analysis for all systems reported in the manuscript, investigating how different relaxation times affect the temperature of trajectories generated with varying prediction horizons $\Delta t$.
Specifically, we focus on three relaxation times, $\tau = 10$, $50$, and $100$ times $\Delta t$, across various prediction horizons.
For each combination of $\tau$ and $\Delta t$, we generate three trajectories by rolling out the MPNNs trained at the respective $\Delta t$ in the NVT ensemble, varying the seed of the thermostat and the initial velocities. 
Each trajectory consists of $100,000$ steps, with the first half truncated for equilibration.\\

\begin{figure*}[b!]
    \centering
    \includegraphics[width=1\linewidth]{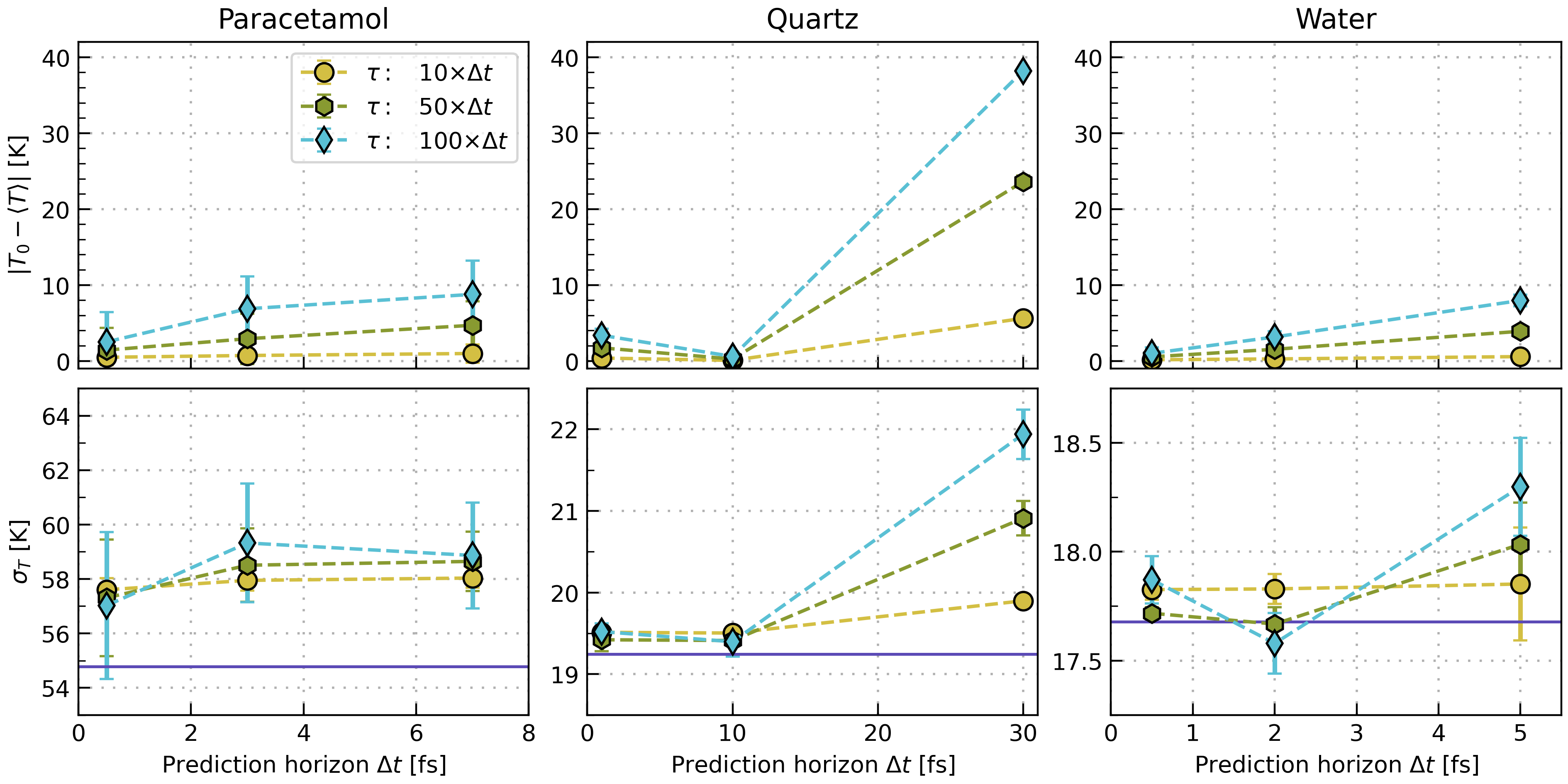}
    \caption{\textbf{Impact of thermostat relaxation time on temperature at different prediction intervals.}
    Top panels show the absolute deviation from the target temperature $T_0=300$~K for the three systems, while bottom panels show the temperature standard deviation, $\sigma_T$, computed from the trajectory.
    The purple lines correspond to the theoretical expectation for $\sigma_T$ from equation~\ref{eq:si_sigma_t}.
    Errors bars represent twice the standard deviation across three different seeds.
    The dashed lines serve as guides to the eye.
    }
    \label{si:fig_thermo_tau}
\end{figure*}

In addition to the deviation of the ensemble average temperature, $\langle T \rangle$, from the target temperature $T_0 = 300$~K, we also evaluate the standard deviation of the temperature, $\sigma_T$, which is expected to follow the relationship \cite{frenkel_understanding_2002,hickman_temperature_2016}:
\begin{equation}
\sigma_T = \sqrt{\frac{2}{3N}}T_0
\label{eq:si_sigma_t}
\end{equation}
where $N$ is the number of atoms in the system.
An overview of these properties is provided in Fig.~\ref{si:fig_thermo_tau}.
While the thermostat achieves the correct target temperature and fluctuations independent of the chosen relaxation time for short prediction horizons, significant differences emerge when $\Delta t$ is increased.
The weakest coupling ($\tau = 100 \times \Delta t$) fails to maintain the target temperature for large prediction horizons, with deviations up to $\approx 15$~K for paracetamol and $\approx 40$~K for quartz, respectively.
While setting $\tau=50\times \Delta t$ reduces the error, only the strongest thermostat performs satisfactory with deviations $\leq 1$~K, except for $\alpha$-quartz using  $\Delta t=30$~fs, which requires an even lower $\tau$.
We observe the same trends in temperature fluctuations based on the standard deviation, where for large prediction horizons, a weak thermostat introduces larger errors.
Beyond affecting ensemble averages and properties, in the course of this work we observed that excessive deviations in the target temperature can induce instabilities.
All this indicates that significantly smaller relative relaxation times are required when using large prediction horizons to accurately reproduce the NVT ensemble.

% We find that the weakest investigated coupling corresponding to $\tau=100\times \Delta t$ struggles to maintain the correct temperature culminating in deviations of up to $10$~K for paracetamol even for small timesteps and even up to $30$~K for $\alpha$-quartz run with $\Delta t = 20$~fs.

% Across all systems and prediction intervals, we find that the weakest coupling corresponding to $\tau=100\times \Delta t$ does not sufficiently 

\subsection{Forecasting at Constant Energy}\label{si:sec_nve}

\begin{figure*}[b!]
    \centering
    \includegraphics[width=0.9\linewidth]{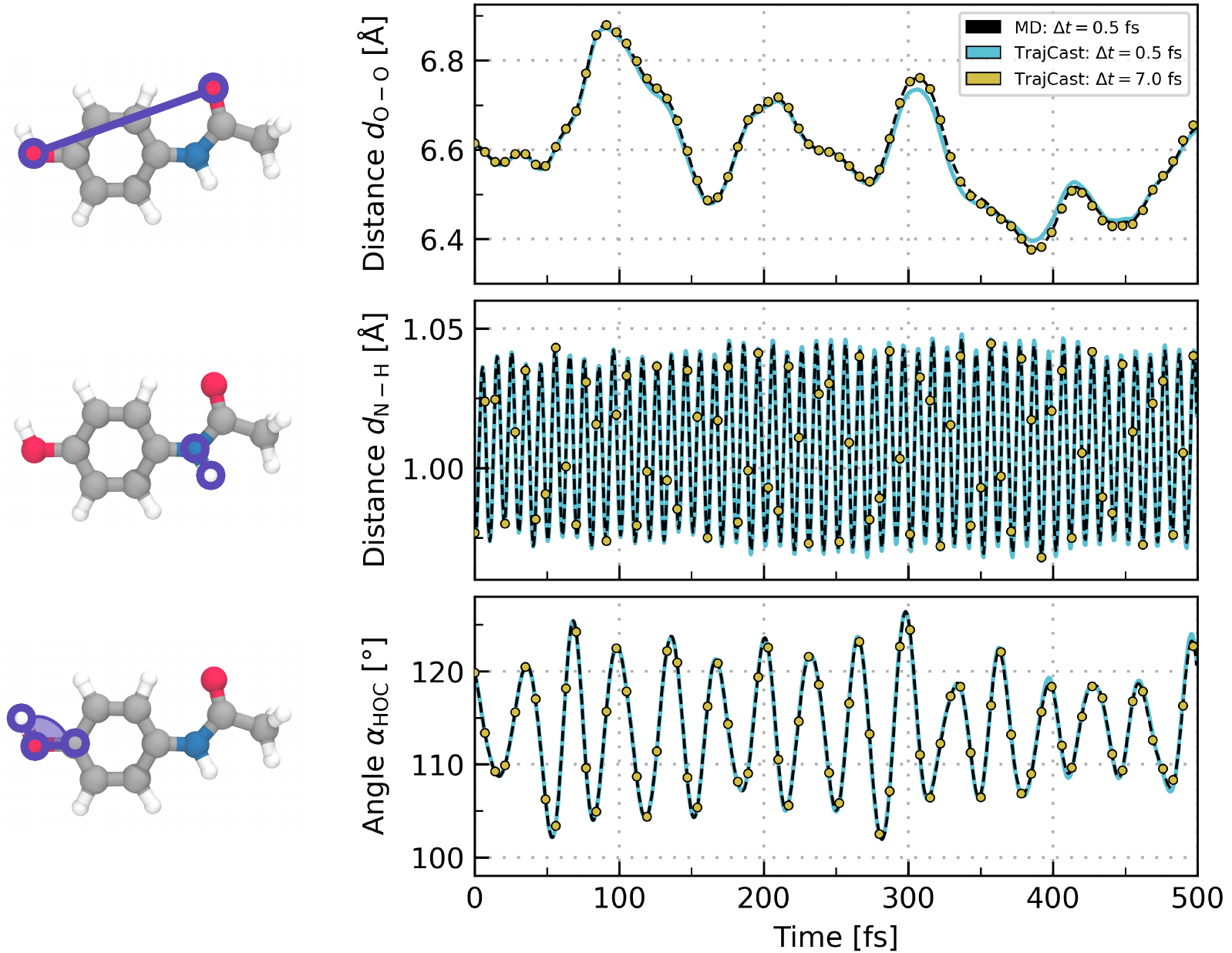}
    \caption{
    \textbf{Comparison of \textit{TrajCast} and MD trajectories of paracetamol in the NVE ensemble.}
    The MD trajectories were generated using a time-step of 0.5~fs and we compare to \textit{TrajCast} roll-out based on prediction horizons of 0.5~fs and 7.0~fs, corresponding to MTMs of 1 and 14, respectively.
    The molecular snapshots on the left highlight different features such as interatomic distances and angles which are used to validate the model performance.
    The right panels show the respective metrics as a function of time of up to 500~fs corresponding to 1,000 MD steps.
    }
    \label{si:fig_nve}
\end{figure*}

While the CSVR thermostat enables \textit{TrajCast} to generate trajectories at constant temperature within the canonical ensemble, the MPNNs driving our framework mimic the system's temporal evolution in the NVE ensemble.
Specifically, we trained these models to predict the new state after multiple steps of numerical integration of the equations of motion using the velocity Verlet algorithm.
Although velocity rescaling stabilizes the roll-out, here, we show that our models can also generate stable trajectories without noise injection, closely following the MD reference.
 To this end, we autoregressively generate trajectories without a thermostat using MPNNs trained to paracetamol with prediction horizons of $\Delta t = 0.5$~fs and $\Delta t = 7.0$~fs, corresponding to MD time-step multiples (MTM) of 1 and 14, respectively, and compare them to a MD NVE trajectory started from the same state.
Each trajectory was generated over 1,000 steps.
 Unlike benchmarking in the NVT ensemble, the lack of non-deterministic rescaling allows for a direct comparison of the trajectories based on the divergence of atomic positions or velocities.
Similar to \cite{winkler_high-fidelity_2022}, in Fig.~\ref{si:fig_nve}, we compare specific interatomic distances and angles of paracetamol as a function of time.
To resolve the model performance across different frequency ranges, we pick the oxygen-oxygen and nitrogen-hydrogen distances as well as the hydroxyl group angle.
Across all metrics we find that both MPNNs achieve almost perfect agreement across the entire range of 500~fs.
Despite these very encouraging results, however, we note that larger errors and instabilities could occur at longer timescales.

\subsection{Forecasting Efficiency}\label{si:sec_efficiency}
Here, we evaluate the computational efficiency of our approach by measuring the number of steps and estimated trajectory length achievable per day across different systems.
Specifically, we perform forecasting in the NVT ensemble for all three systems over 1,000 steps using the models from the experiments reported in the manuscript.
To assess only the efficiency of forward passes and velocity rescaling, no frames are saved.
For the condensed matter systems of quartz and liquid water, we additionally report these timings for $2\times2\times2$ and $3\times3\times3$ supercells.
Throughout all systems, we made sure that the trajectories were stable.
Timings were measured on a NVIDIA V100-SXM2-32GB GPU for the isolated atom and up to $2\times2\times2$ supercells, while larger systems were benchmarked on an NVIDIA A100-SXM4-40GB GPU due to memory requirements for the largest water system.
An overview of the system and timing details is provided in Table~\ref{tab:timings}, where we report the fastest run across multiple trials on different devices of the same type.
Any observed deviations from the values reported  were within approximately 20 seconds.
Based on the time for 1,000 steps and the prediction horizon, we extrapolate to estimate the approximate length of a trajectory that \textit{TrajCast} can generate for each system within one day.\\

Interestingly, the generation of the the trajectory for the molecule comprising 20 atoms only is on the same order of magnitude as smallest quartz and water systems being almost one order of magnitude larger.
Besides potential GPU inefficiencies at this small system sizes, this may also stem from angular momentum conservation, which is incorporated only into the MPNN trained for the gas-phase molecule.
Future work could explore this by training a model on a peptide or protein in vacuum.
The deviations between quartz and water, conversely, originate from the different cut-offs of 4.5~$\text{\AA}$ and 6.0~\AA, respectively.
Due to the large prediction horizon for quartz, \textit{TrajCast} can generate nearly 20 ns of trajectory for over 4,000 atoms, while for water, it achieves just over 1 ns per day for a system of about 5,000 atoms.
We note, however, that these values are merely rough guidelines and were obtained under testing environments 
avoiding GPU-to-CPU transfers and disk writing.

% which are then transferred to NVT by the thermostat.

\begin{table}[h!]
    \caption{
    \textbf{Computational efficiency of forecasting across chemical systems and sizes.}
    These benchmarks were obtained by generating 1,000 steps of a trajectory in the NVT ensemble.
    To purely estimate the inference time, no intermittent atomic configurations were saved.    
    Forecasts for the largest supercells of quartz and water were performed on a NVIDIA A100-SXM4-40GB GPU, while all other systems were assessed on a NVIDIA V100-SXM2-32GB GPU.
    }
    \label{tab:timings}
    \resizebox{\textwidth}{!}{
        \begin{tabular}{lccccc}
             \toprule
             \midrule
        System & Supercell & $N_\mathrm{atoms}$& Prediction horizon $\Delta t$ [fs]&  Time per 1,000 steps [s] & Length per day [ns]
             \\ 
             \midrule
             \midrule
         Paracetamol & $1\times 1\times 1$ & 20 & 7.0 & 57.65 & 10.49 \\
         \midrule
         \multirow{3}{*}{Quartz} & $1\times 1\times 1$ & 162 & \multirow{3}{*}{30.0}& 53.75 &  48.22 \\
          & $2\times 2\times 2$ & 1,296 & & 75.16 & 34.49\\
           & $3\times 3\times 3$ & 4,374 & & 145.68 & 17.79 \\
           \midrule
                 \multirow{3}{*}{Water} & $1\times 1\times 1$ & 192 & \multirow{3}{*}{5.0}& 62.02 & 6.97 \\
          & $2\times 2\times 2$ & 1,536 & & 162.85 & 2.65 \\
           & $3\times 3\times 3$ & 5,184 &  & 371.38 & 1.16\\
         \midrule
            \bottomrule
        \end{tabular}}
\end{table}

% While we are generally interested properties measured at constant temperature rather than temperature, it is important to showcase that \textit{TrajCast}, or more specifically, the trained MPNNs are able to accurately predict the temporal evolution of the system in the NVE ensemble.